\documentclass[10pt,journal,twocolumn,twoside]{autart} 
\usepackage{graphicx}
\usepackage{amssymb}
\usepackage{amsmath}
\usepackage{algorithm}
\usepackage{algpseudocode}
\usepackage{subcaption}
\usepackage{hyperref}
\usepackage{comment}
\usepackage{color}
\usepackage{cite}

\usepackage{scalerel}

\newcommand{\R}{{\mathbb{R}}}
\newcommand{\Z}{{\mathbb{Z}}}
\newcommand{\Xf}{{\mathbb{X}_{\mathrm{f}}}}
\newcommand{\D}{\mathbb{D}}
\newcommand{\ka}{\kappa(x,z,v)}
\newcommand{\I}[1]{\mathbb{I}_{[#1]}}

\begin{document}

\begin{frontmatter} 
\title{Robust adaptive MPC using control contraction metrics}
\author{Andr\'as Sasfi$^1$}, %
\author{Melanie N. Zeilinger$^2$}, %
\author{Johannes K\"ohler$^2$}%
\address{$^1$Automatic Control Laboratory, ETH Z\"urich, Z\"urich CH-8092, Switzerland}
\address{$^2$Institute for Dynamic Systems and Control, ETH Z\"urich, Z\"urich CH-8092, Switzerland}
\date{\copyright 2023 Elsevier Ltd. All rights reserved.}

\begin{abstract}
We present a robust adaptive model predictive control (MPC) framework for nonlinear continuous-time systems with bounded parametric uncertainty and additive disturbance.
We utilize general control contraction metrics (CCMs) to parameterize a homothetic tube around a nominal prediction that contains all uncertain trajectories. 
Furthermore, we incorporate model adaptation using set-membership estimation.
As a result, the proposed MPC formulation is applicable to a large class of nonlinear systems, reduces conservatism during online operation, and guarantees robust constraint satisfaction and convergence to a neighborhood of the desired setpoint. 
One of the main technical contributions is the derivation of corresponding tube dynamics based on CCMs that account for the state and input dependent nature of the model mismatch. 
Furthermore, we online optimize over the nominal parameter, which enables general set-membership updates for the parametric uncertainty in the MPC. 
Benefits of the proposed homothetic tube MPC and online adaptation are demonstrated using a numerical example involving a planar quadrotor.  
\end{abstract}
\end{frontmatter} 
\section{Introduction} 
Some of the most challenging control problems are characterized by nonlinear uncertain dynamics and safety critical constraints. 
Optimization-based control techniques, such as model predictive control (MPC)~\cite{rawlings2017model}, can offer high performance control for nonlinear systems, while respecting general state and input constraints. 
However, the desired closed-loop properties (stability, performance, constraint satisfaction) are in general not preserved under model uncertainty. 
In this paper, we develop a robust adaptive MPC framework that robustly ensures such closed-loop properties for a large class of uncertain nonlinear systems, while incorporating online model updates to reduce conservatism. 
This is achieved by introducing a novel \textit{robust} MPC design, including a general method to incorporate \textit{adaptive} model updates. 

\subsubsection*{Robust MPC methods for nonlinear uncertain systems}
Robust MPC (RMPC) methods typically ensure constraint satisfaction despite model uncertainty by predicting a tube/funnel over the prediction horizon that contains all uncertain trajectories.
There exists a wide range of nonlinear RMPC approaches in the literature \cite{lucia2015robust,villanueva2017robust,houska2019robust,leeman2023robust,zanelli2021zero}, which vary strongly in their conservatism or computational complexity.
For example, System Level Synthesis~\cite{leeman2023robust} allows to jointly optimize over a linear feedback and optimization over cross sections of the tube results in great flexibility~\cite{villanueva2017robust,houska2019robust,zanelli2021zero}. However, these methods can result in a drastic increase of the computational complexity that may impede real-time implementation.
The trade-off between complexity and conservatism is well-understood in case of linear systems (cf., e.g.,~\cite{kouvaritakis2016model,chen2023robust}).
In this work, we focus on the design of nonlinear robust MPC methods with small computational complexity, ideally close to nominal MPC. 
In case of linear systems, the tube-based RMPC schemes in~\cite{mayne2005robust,rakovic2012homothetic,parsi2022scalable} achieve this by performing more expensive computations related to
reachability analysis and invariant sets offline. 
This results in a parameterized tube, e.g., in the form of a polytope or ellipsoid, which is simple to predict over the prediction
horizon.
Therein, computational efficiency is ensured by performing more expensive computations related to reachability analysis and invariant sets offline, resulting in a parameterized tube, e.g., in the form of a polytope or ellipsoid, that is simple to predict over the prediction horizon.
Correspondingly, utilizing similar concepts for nonlinear systems requires an offline computation of (parameterized) robust positive invariant sets, which can be done using simple ellipsoidal sets~\cite{yu2013tube,wabersich2021nonlinear}, general incremental Lyapunov functions~\cite{bayer2013discrete,Koehler2020Robust}, and more recently control contraction metrics (CCMs)~\cite{singh2017robust,singh2019robust,zhao2021tube}. 
These invariant sets can then be used as a \textit{rigid} tube around a nominal prediction to obtain a simple nonlinear RMPC scheme~\cite{yu2013tube,wabersich2021nonlinear,bayer2013discrete,singh2017robust,singh2019robust,zhao2021tube}, compare also~\cite{mayne2005robust}. 
However, such a rigid tube is conservative if the model mismatch depends on the state or input, e.g., due to parametric uncertainty. 
Such effects can be better captured using a \textit{homothetic} tube~\cite{lopez2019adaptive,Koehler2020Robust,koehler2021output,kohler2020RAMPC,rakovic2022homothetic}, i.e., the tube scaling depends on the nominal trajectory, compare also~\cite{rakovic2012homothetic,lorenzen2019robust}.
In this paper, we extend and unify these concepts by deriving a homothetic tube MPC framework for nonlinear uncertain systems based on general CCMs. 

\subsubsection*{Parameter updates in robust \textit{adaptive} MPC}
Model mismatch is often partially due to inaccurate system knowledge and performance may be enhanced by incorporating online model updates in RMPC frameworks~\cite{hewing2020learning}. 
In case of linearly parameterized models with bounded disturbance, we can efficiently update parameter sets using set-membership estimation~\cite{lorenzen2019robust,kohler2020RAMPC} or Lyapunov-type arguments~\cite{Adetola2009adaptive,guay2015robust}. 
Especially, utilizing set-membership estimation in RMPC schemes to reduce uncertainty is well-established for linear systems~\cite{lorenzen2019robust,lu2021robust,tanaskovic2014adaptive,lu2023robust,parsi2022dual}. However, since most (nonlinear) RMPC frameworks explicitly depend on a nominal trajectory with some nominal parameter, most resulting RAMPC schemes (cf. \cite{kohler2019linear,didier2021robust,Adetola2009adaptive,guay2015robust,lopez2019adaptive,kohler2020RAMPC}) require restrictive additional assumptions to ensure recursive feasibility under model updates, such as the restriction of the parameter set to balls/hypercubes instead of general ellipsoids/polytopes. 
In this paper, we show how to incorporate general parameter updates in nonlinear RMPC schemes without posing such additional restrictions, resulting in a flexible RAMPC scheme. 

\subsubsection*{Contribution}
The theoretical contribution of this work is twofold. First, we propose a novel robust MPC framework for nonlinear continuous-time systems by deriving a homothetic tube propagation for general CCMs.
Second, we present a method that allows us to naturally incorporate online model updates, resulting in a RAMPC framework.

To formulate the proposed RMPC scheme, we first extend existing results on (robust) CCMs (cf. \cite{lohmiller1998contraction,manchester2017control,singh2017robust,wang2020virtual,zhao2021tube}) by deriving a differential equation for the tube scaling, resulting in a homothetic tube that contains all possible nonlinear uncertain trajectories. 
In particular, the derived tube dynamics directly use state and input dependent bounds on the model mismatch, thus avoiding the conservatism of constant bounds~\cite{singh2017robust,zhao2021tube}. 
To the best knowledge of the authors, this is also the first result on tube MPC schemes that rigorously shows that the CCM conditions are only needed on the constraint set, without imposing additional assumptions. 
Compared to existing homothetic tube MPC formulations (cf.~\cite{lopez2019adaptive,Koehler2020Robust,koehler2021output,rakovic2022homothetic}), the reliance on established CCM synthesis tools (cf.~\cite{manchester2017control,singh2017robust,zhao2021tube}) ensures the practical applicability to a large class of uncertain nonlinear systems. 

We incorporate online model updates by using set-membership estimation for the parametric uncertainty. 
In contrast to existing approaches, we propose to additionally optimize over the nominal parameter that characterizes the nominal trajectory. 
This seemingly trivial modification retains the desired flexibility of the model updates without requiring the additional restrictions needed in related work~\cite{kohler2019linear,didier2021robust,Adetola2009adaptive,guay2015robust,lopez2019adaptive,kohler2020RAMPC}. 
As a result, the derived RAMPC scheme can also directly track steady-states which explicitly depend on the unknown parameter. 

Overall, the resulting homothetic tube RAMPC framework guarantees recursive feasibility, constraint satisfaction, and convergence to a neighborhood of the desired steady-state.
We demonstrate the benefits of the homothetic tube MPC and online model updates using the planar quadrotor example from~\cite{zhao2021tube}.

\subsubsection*{Outline} 
Section~\ref{sec:prelim} introduces the problem setup and outlines the general RAMPC approach.  
Section~\ref{sec:main} presents the main results, including the novel homothetic tube propagation based on CCMs, the proposed RAMPC scheme, and the theoretical analysis. 
Finally, Section~\ref{sec:example} presents a numerical example and Section~\ref{sec:conclusion} provides a conclusion. 

\subsubsection*{Notation}
The set of non-negative real numbers is $\R_{\geq 0}$, while $\I{a,b}$ is the set of integers in the interval $[a,b]$. The $i$-th element of a vector $x$ is denoted by $[x]_i$. Furthermore, $[X]_{:,k}$ is the $k$-th column of $X$.
The zero matrix and the identity matrix of appropriate dimensions are denoted by $0$ and $I$, respectively.  
We use the notation $\langle A \rangle = A + A^\top$ for square matrices $A\in\R^{n\times n}$. 
The set of positive definite matrices of size $n \times n$ is $\mathbb{S}^+_n$, and $A \preceq B$ indicates that $B-A$ is positive semi-definite. 
The quadratic norm of a vector $x$ w.r.t. a positive definite matrix $M$ is denoted by $\|x\|_{M} := \sqrt{x^\top M x}$. 
The Cholesky decomposition of a positive definite matrix is denoted by $\left (M^{\frac{1}{2}} \right)^\top M^{\frac{1}{2}} = M$, yielding $\left\|x\right\|_M = \left\|M^{\frac{1}{2}}x \right\|$. Throughout the paper, we indicate the time dependence by $x(t)$ for actual trajectories or $z_{t|t_k}$ for predicted trajectories at time $t_k$. For a continuously differentiable function $f:\R^a \times \R^b \to \R^c$, $f(x,u)$, the partial and total derivative w.r.t. $x$ evaluated at some point $(x,u) = (z,v)$ is defined as $\left.\dfrac{\partial f}{\partial x}\right|_{(z,v)} \in \R^{c \times a}$ and $\left.\dfrac{\mathrm{d}f}{\mathrm{d} x}\right|_{(z,v)}$, respectively. Similarly, we abbreviate the total time derivative by $\dot{f}(x,u):=\left.\dfrac{\mathrm{d}f}{\mathrm{d}t}\right|_{(x,u)}$.

\section{Preliminaries} \label{sec:prelim}
We first present the considered problem setup (Sec.~\ref{sec:setup}). 
Then, we outline how the problem is addressed using the proposed homothetic tube RAMPC scheme (Sec.~\ref{sec:problem-statement}). 

\subsection{Problem setup} \label{sec:setup}
We consider a continuous-time nonlinear system
\begin{align} \label{eq:system}
\dot{x}(t)=f_\mathrm{w}(x(t),u(t),\theta,d(t)), \quad x(0) = x_0, \quad t \geq 0,
\end{align}
with state $x(t)\in\mathbb{R}^n$, initial state $x_0\in\mathbb{R}^n$, input $u(t)\in\mathbb{R}^m$, disturbance $d(t) \in \mathbb{R}^q$, and constant model parameter $\theta \in \R^p$. 
The disturbance $d(t)$ and the model parameter $\theta$ are unknown and satisfy a known bound $d(t)\in \D$, $t\geq0$, $\theta \in \Theta_0$ with polytopes $\Theta_0 \subseteq \R^p$, $\mathbb{D} \subseteq \mathbb{R}^q$ and $0\in\mathbb{D}$, and the state $x(t)$, $t\geq 0$ can be measured. 
Furthermore, we assume that the dynamics $f_{\mathrm{w}}$ are affine in $u$, $d$, $\theta$, i.e., we have
\begin{align*}
f_\mathrm{w}(x,u,\theta,d)  = f(x) + B(x)u + G(x,u)\theta+{E(x)}d,\nonumber
\end{align*}
with $G$ affine in $u$, i.e., $f_{\mathrm{w}}$ is affine in $u$, $d$ and $\theta$, respectively.
Throughout the paper, we assume that $f_\mathrm{w}$ is continuously differentiable and Lipschitz continuous
and $u(t)$ and $d(t)$ are piecewise continuous, which implies existence of a unique solution $x(\cdot)$ to the system \eqref{eq:system}, see \cite[Thm. 3.2]{khalil2002nonlinear}.\footnote{Similarly to \eqref{eq:constraint-tight1}--\eqref{eq:c_j-def}, one can show that the feedback $\kappa(x,z,v)$ in Proposition~\ref{prop:V-delta} is Lipschitz continuous w.r.t. $x$. Hence, existence and uniqueness of the solution to \eqref{eq:system} remain valid under the feedback $\kappa$.}
The system is subject to state and input constraints 
\begin{equation} \label{eq:constr}
(x(t),u(t))\in\Z, \quad t \geq 0,
\end{equation}
where $\Z$ is a compact set defined by
\begin{equation} \label{eq:Z_set}
\Z:=\left\{(x,u)\in\mathbb{R}^{n+m}|h_j(x,u)\leq 0,~j \in \I{1,r}\right\}.
\end{equation}
We denote the projection of $\Z$ on $x$ by $\Z_\mathrm{x}$.

The overall goal can be summarized in the following conceptual infinite-horizon optimal control problem:
\begin{equation} \label{eq:conceptual}
    \begin{split}
    &\min_{\pi(\cdot)} \; \max_{d(\cdot) \in \D,~\theta \in \Theta_0} \mathcal{J}_\infty(x(0),\pi(\cdot),d(\cdot),\theta) \\
    \textnormal{s.t.} \quad & u(t) = \pi(x(\cdot),t), \\
    &(x(t),u(t)) \in \Z,\\
    &\dot{x}(t) = f_\mathrm{w}(x(t),u(t),\theta,d(t)), \quad t \geq 0.
    \end{split}
\end{equation}
The goal is to find a causal policy $\pi$, that minimizes some cost $\mathcal{J}_\infty$ under the worst-case disturbance and parameter, while robustly satisfying the state and input constraints. 

\subsection{General RAMPC approach} \label{sec:problem-statement}
Problem~\eqref{eq:conceptual} is intractable in this form due to the optimization over a general policy $\pi$ that depends on all past data and the difficulty to compute the exact (disturbance) reachable set for nonlinear systems. 
In the following, we provide a high-level description of how we 
compute a feasible solution to Problem~\eqref{eq:conceptual} by using a finitely parametrized policy $\pi$ and an over-approximation of the possible state trajectories $x(t) \in \mathbb{T}_t$, where $\mathbb{T}_t$ is called a tube.
As common in tube-based robust MPC formulations, this is achieved by defining a nominal state and input trajectory $z_t$, $v_t$ based on some nominal parameter $\overline{\theta}\in\Theta_0$ with the nominal dynamics $\dot{z}_t = \overline{f}(z_t,v_t,\overline{\theta}) := f_\mathrm{w}(z_t,v_t,\overline{\theta},0)$. 
The policy is then given as a feedback law $\kappa:\R^n \times \Z \rightarrow \R^m$ parametrized by the nominal trajectories, i.e., $\pi(x(\cdot),t) = \kappa(x(t),z_t,v_t)$. Furthermore, the tube is parametrized as
\begin{align} \label{eq:tube-def}
    \mathbb{T}_t = \{x \in \R^n|V_\delta(x,z_t) \leq \delta_t \},\quad t\geq 0,
\end{align} where $V_\delta: \R^n \times \R^n \rightarrow \R_{\geq 0}$ is a later specified Lyapunov-type function and $\delta_t \geq 0$ is a suitable scaling. 
Given this parametrization, we can provide a feasible solution to Problem~\eqref{eq:conceptual} with
\begin{subequations}
\label{eq:tube_conceptual}
\begin{align}
\label{eq:tube_conceptual_1}
&\dot{z}_t=\overline{f}(z_t,v_t,\overline{\theta}),\quad \dot{\delta}_t=f_\delta(z_t,v_t,\delta_t,\overline{\theta},\Theta_0),\\
\label{eq:tube_conceptual_2}
&(x(t),\kappa(x(t),z_t,v_t))\in\mathbb{Z},\quad \forall x(t)\in\mathbb{T}_t,\quad t\geq 0,
\end{align}
\end{subequations}
where $f_\delta$ corresponds to a later derived dynamics for the tube scaling (cf. Sec.~\ref{sec:tube-design}) ensuring $x(t)\in\mathbb{T}_t$, $t\geq 0$. 
Since the parametrization~\eqref{eq:tube-def} contains a (not necessarily linear) translation $z_t$ and a scaling $\delta_t$, 
we refer to it as a homothetic tube MPC due to its similarity to existing approaches for linear systems~\cite{lorenzen2019robust,rakovic2012homothetic}. 
In particular, in case $V_\delta$ is a polytopic or quadratic function of the error $x-z_t$, we recover the homothetic tube parametrization in~\cite{didier2021robust,kohler2019linear,lorenzen2019robust,rakovic2012homothetic,parsi2022scalable,rakovic2022homothetic}. 
As one of the main technical contributions, in Sections~\ref{sec:CCM}--\ref{sec:tube-design} we show how suitable functions $\kappa$, $V_\delta$ can be constructed offline using CCMs and derive dynamics for the tube scaling $\delta_t$.

For simplicity, we assume that the cost $\mathcal{J}_\infty$ in~\eqref{eq:conceptual} corresponds to a quadratic tracking stage cost of the form
\begin{equation} \label{eq:stage-cost-def}
    \ell(z,v,\overline{\theta}) = \|z-z_\mathrm{ref}(\overline{\theta})\|_Q^2 + \|v - v_\mathrm{ref}(\overline{\theta})\|_R^2,
\end{equation}
with positive definite matrices $Q \in \R^{n \times n}$, $R \in \R^{m \times m}$ and a feasible steady-state and input $(z_{\mathrm{ref}}(\overline{\theta}),v_{\mathrm{ref}}(\overline{\theta}))\in \Z$, $\overline{f}(z_{\mathrm{ref}}(\overline{\theta}),v_{\mathrm{ref}}(\overline{\theta}),\overline{\theta})=0$, $\overline{\theta}\in\Theta_0$, which may depend (continuously) on the nominal parameter $\overline{\theta}$ (cf. also Sec.~\ref{sec:discussion}). 
To approximate the infinite-horizon problem~\eqref{eq:conceptual}, we repeatedly solve a finite-horizon problem at every sampling time $t_k = kT_\mathrm{s}$, $k \in \mathbb{I}_{\geq0}$ with sampling period $T_\mathrm{s} > 0$, using the standard receding horizon principle from MPC~\cite{rawlings2017model}. 
We incorporate model adaptation by updating the set $\Theta_0$ online at every sampling time $t_k$ using set-membership estimation.

\section{Robust adaptive MPC framework} \label{sec:main}
This section derives the proposed RAMPC framework, which is the main result of the paper. First, we use CCMs to construct the homothetic tube (Sec.~\ref{sec:CCM}--\ref{sec:tube-design}). Next, we present how the initial parameter set $\Theta_0$ can be updated using set-membership estimation (Sec.~\ref{sec:set-update}), followed by the design of the terminal ingredients (Sec.~\ref{sec:terminal}). Then, the overall algorithm, including the MPC problem, is described and the theoretical properties are derived (Sec.~\ref{sec:framework}). Finally, we provide some discussion, including a qualitative comparison to existing approaches (Sec.~\ref{sec:discussion}).

\subsection{Control contraction metrics} \label{sec:CCM}
In the following, we show that CCMs provide a natural way to derive the ingredients needed in the homothetic tube design ($V_\delta,\kappa,f_\delta$). In particular, CCMs use the Jacobian of the dynamics $f_{\mathrm{w}}$ to synthesize controllers ensuring (incremental) stability for the nonlinear dynamics~\cite{manchester2017control,singh2017robust,singh2019robust,wang2020virtual,zhao2021tube}. 
\begin{assum} \label{ass:CCM}
There exists a smooth function $M: \R^n \to \mathbb{S}^+_n$, a continuous function $K:\R^n \to \R^{m \times n}$, a contraction rate $\rho_\mathrm{c} > 0$, and matrices $\underline{M},~\overline{M} \in \mathbb{S}^+_n$, such that for all $(x,u) \in \Z$, and all $\theta \in \Theta_0$, $d \in \D$, the following properties hold:
\begin{subequations}
\label{eq:CCM} 
\begin{align}
    \dot{M}(x)+ \langle M(x) A_\mathrm{cl}(x,u,\theta,d) \rangle \preceq & -2\rho_\mathrm{c} M(x), \label{eq:CCM-contraction} \\
    \underline{M} \preceq M(x) \preceq & \overline{M}, \label{eq:CCM-bounded-M} 
\end{align}
\end{subequations}
with $\dot{x} = f_\mathrm{w}(x,u,\theta,d)$, and
\begin{equation*}
     A_\mathrm{cl}(x,u,\theta,d) := \left.\dfrac{\partial f_\mathrm{w}}{\partial x}\right|_{(x,u,\theta,d)} + \left.\dfrac{\partial f_\mathrm{w}}{\partial u}\right|_{(x,u,\theta,d)}K(x).
\end{equation*}
\end{assum}
Conditions \eqref{eq:CCM-contraction}--\eqref{eq:CCM-bounded-M} imply stability of the linearized dynamics with the differential feedback $K$ using the contraction metric $M$. By integrating this differential form we can obtain stability properties for the nonlinear system. To this end, we denote the set of piece-wise smooth curves $\gamma : [0,1] \rightarrow \R^n$ with $\gamma(0) = z$ and $\gamma(1) = x$ by  $\Gamma(z,x)$, and define $\gamma_{\mathrm{s}}(s) := \left.\dfrac{\partial \gamma}{\partial s}\right|_s$. Given a CCM satisfying Assumption~\ref{ass:CCM}, we define a corresponding incremental Lyapunov function $V_\delta$ (cf.~\cite{manchester2017control}) as follows:
\begin{equation} \label{eq:V-delta-def}
V_\delta(x,z) := \min_{\gamma \in \Gamma(z,x)} \int_{0}^{1} \|\gamma_{\mathrm{s}}(s)\|_{M(\gamma(s))} \mathrm{d}s.
\end{equation}
A minimizer is called a geodesic $\gamma^\star$ (with $\gamma_{\mathrm{s}}^\star$), which we assume to exist (cf.~\cite{manchester2017control} for corresponding conditions). Integration of the local feedback $K(x)$ along the geodesic $\gamma^\star$ yields the input $\gamma^{\mathrm{u}}$, and the feedback law $\kappa$:
\begin{align} \label{eq:kappa-def}
\begin{split}
    \gamma^{\mathrm{u}}(s) :=& v + \int_0^{s} K(\gamma^\star(\tilde{s}))\gamma_{\mathrm{s}}^\star(\tilde{s}) \mathrm{d}\tilde{s}, \\
    \kappa(x,z,v) :=& \gamma^{\mathrm{u}}(1). 
\end{split}
\end{align} 
The following proposition summarizes the properties of $V_\delta$ and $\kappa$. 
\begin{prop} \label{prop:V-delta}
Suppose Assumption~\ref{ass:CCM} holds. Consider some $(x,z,v)\in \R^n \times \R^n \times \R^m$ such that for all $s\in[0,1]$ it holds that $(\gamma^\star(s),\gamma^{\mathrm{u}}(s))\in\Z$. Then, for all $\theta, \overline{\theta} \in \Theta_0$, and all $d \in \D$, the following inequalities hold:
\begin{subequations}
\begin{align}
    \|x-z\|_{\underline{M}} \leq& V_\delta(x,z) \leq \|x-z\|_{\overline{M}}, \label{eq:Vd-bound} \\
	\dot{V_\delta}(x,z) =& \left.\dfrac{\partial V_\delta}{\partial x}\right|_{(x,z)} \dot{x}
	+  \left.\dfrac{\partial V_\delta}{\partial z}\right|_{(x,z)} \dot{z} \label{eq:Vd-contraction}\\
	\leq& - \left(\rho_\mathrm{c} - L_\D - \sum_{k=1}^{p}L_{\mathrm{G},k}\left|[\theta - \overline{\theta}]_k\right| \right) V_\delta(x,z) \nonumber\\
	&+ \|G(z,v) (\theta-\overline{\theta}) + E(z)d\|_{M(z)}, \nonumber
\end{align}
with $\dot{x} = f_\mathrm{w}(x,\ka,\theta,d)$, $\dot{z} = \overline{f}(z,v,\overline{\theta})$, $V_\delta$, $\kappa$ according to
Equations~\eqref{eq:V-delta-def}, \eqref{eq:kappa-def}, and constants $L_{\mathrm{G},k} \geq 0$, $k \in \I{1,p}$, $L_\D \geq 0$ according to \eqref{eq:app-L-G} and \eqref{eq:app-L-D} in the appendix.
\end{subequations}
\end{prop}
The proof of this proposition is given in Appendix~\ref{app:proof-V-delta}. In the nominal setting, i.e., $\mathbb{D}=\{0\}$, $\theta=\overline{\theta}$, Proposition~\ref{prop:V-delta} ensures that $V_\delta$ is an incremental Lyapunov function for $\dot{x}=\overline{f}(x,u,\overline{\theta})$ with feedback $\kappa(x,z,v)$, analogous to~\cite{manchester2017control}. 
The effect of the disturbance $d$ and the parametric uncertainty $\theta-\overline{\theta}$ is suitably bounded with the constants $L_\D$, $L_{\mathrm{G},k}$ modifying the contraction rate $\rho_\mathrm{c}$, and the model mismatch $\|G(z,v) (\theta-\overline{\theta}) + E(z)d\|_{M(z)}$ evaluated at the nominal trajectory. 
In the next section, we demonstrate that the condition $(\gamma^\star(s),\gamma^{\mathrm{u}}(s)) \in \Z$, $s \in [0,1]$ is intrinsically satisfied by the proposed tube MPC and derive the tube dynamics based on  Inequality~\eqref{eq:Vd-contraction}.

\begin{rem} \label{rem:CCM-computation}
The offline computation of matrices $M$, $K$ satisfying Assumption~\ref{ass:CCM} can be cast as linear matrix inequalities (LMIs), or a sum of squares (SOS) problem, using the convex re-parametrization $W=M^{-1}$ and $Y = KW$, compare~\cite{manchester2017control,singh2019robust,zhao2021tube,wang2020virtual} for details.  
The geodesic $\gamma^\star$ is used in Equations~\eqref{eq:V-delta-def} and \eqref{eq:kappa-def} and can be computed using the Chebyshev global pseudospectral method with a suitable discretization~\cite{leung2017nonlinear}. 
\end{rem}

\begin{rem}
In contrast to Proposition~\ref{prop:V-delta}, most of the CCM literature only considers a nominal setting~\cite{lohmiller1998contraction,manchester2017control}. 
Closer to our work, the disturbance $d$ is explicitly considered in the CCM conditions in~\cite{zhao2021tube}. 
While~\cite{zhao2021tube} results in a robust positive invariant set, we derived a state and input dependent bound, which will be exploited in the homothetic tube MPC formulation.  
We note that the CCM parametrization in Assumption~\ref{ass:CCM} can also be generalized to allow for a dependence of $M$, $K$ on the nominal parameters $\overline{\theta}$, as suggested in~\cite{lopez2021universal} for adaptive control.
\end{rem}

\subsection{Homothetic tube} \label{sec:tube-design}
For the homothetic tube MPC formulation, we need conditions on the tube scaling $\delta$ to ensure that all uncertain trajectories are confined in the tube $\mathbb{T}_t$~\eqref{eq:tube-def} and satisfy the constraints. 
First, the following proposition provides sufficient conditions for the constraint satisfaction inside the tube (cf. Eq.~\eqref{eq:tube_conceptual_2}). 
\begin{prop} \label{prop:c_j}
Suppose Assumption~\ref{ass:CCM} holds. Then, for any $x,z \in \R^n$, $v \in \R^m$ satisfying
\begin{subequations}
  \begin{align}
    &h_j(z,v) + c_j V_\delta(x,z) \leq 0, \quad j \in \I{1,r}, \label{eq:constraint-tight1}\\
    &c_j := \max_{(z,v) \in \Z} \left \| \left(\left.\dfrac{\partial h_j}{\partial x}\right|_{(z,v)} + \left.\dfrac{\partial h_j}{\partial u}\right|_{(z,v)} K(z) \right) M(z)^{-\frac{1}{2}} \right \| \label{eq:c_j-def}
  \end{align}
  it holds
  \begin{align} \label{eq:constraint-sat}
    (\gamma^\star(s),\gamma^\mathrm{u}(s)) \in \Z, \quad s \in [0,1],
  \end{align}
\end{subequations}
with $\gamma^\star$ and $\gamma^\mathrm{u}$ according to Equations \eqref{eq:V-delta-def} and \eqref{eq:kappa-def}.
\end{prop}
\begin{pf}
Note that~\eqref{eq:constraint-tight1} yields $h_j(z,v) \leq 0$, $j \in \I{1,r}$. 
We only consider $c_j>0$, as $c_j=0$ implies $\left. \dfrac{\mathrm{d} h_j}{\mathrm{d} s}\right|_{(\gamma^\star(s),\gamma^\mathrm{u}(s))} = 0$ for all $s \in [0,1]$, and thus $h_j(\gamma^\star(s),\gamma^\mathrm{u}(s)) = h_j(z,v) \leq 0$, $s \in [0,1]$, $j \in \I{1,r}$ follows trivially. For contradiction, suppose that $h_{j}(\gamma^\star(\hat{s}),\gamma^\mathrm{u}(\hat{s})) > 0$ for some $\hat{s} \in (0,1]$, and some $j \in \I{1,r}$. Since $h_{j}(\gamma^\star(0),\gamma^\mathrm{u}(0)) \leq 0$, and $h_{j}(\gamma^\star(s),\gamma^\mathrm{u}(s))$ is continuous w.r.t. $s$, there exists a $\overline{s}\in [0,\hat{s})$, such that $\max_{j \in \I{1,r}} h_j(\gamma^\star(\overline{s}),\gamma^\mathrm{u}(\overline{s})) =0$, and $(\gamma^\star(s),\gamma^\mathrm{u}(s)) \in \Z$, $s \in [0,\overline{s}]$. The gradient theorem yields
\begin{align*}
  &h_j(\gamma^\star(\overline{s}),\gamma^\mathrm{u}(\overline{s})) - h_j(z,v) = \int_0^{\overline{s}} \left.\dfrac{\mathrm{d} h_j}{\mathrm{d} s}\right|_{(\gamma^\star(s),\gamma^\mathrm{u}(s))} \mathrm{d}s \\
  =& \int_0^{\overline{s}} \left (\left.\dfrac{\partial h_j}{\partial x}\right|_{(\gamma^\star(s),\gamma^\mathrm{u}(s))} \right.\\
  & + \left. \left.\dfrac{\partial h_j}{\partial u}\right|_{(\gamma^\star(s), \gamma^\mathrm{u}(s))} K(\gamma^\star(s)) \right) \cdot \gamma_{\mathrm{s}}^\star(s) \mathrm{d} s \\
  \stackrel{\eqref{eq:c_j-def}}{\leq} & c_j \int_0^{\overline{s}} \left \| M(\gamma^\star(s))^{\frac{1}{2}} \gamma^\star_s(s) \right \| \mathrm{d} s  \stackrel{\eqref{eq:V-delta-def}}{<} c_j V_\delta(x,z) \\
  \stackrel{\eqref{eq:constraint-tight1}}{\leq}& -h_j(z,v), \quad j \in \I{1,r},
\end{align*}
which contradicts $\max_{j \in \I{1,r}} h_j(\gamma^\star(\overline{s}),\gamma^\mathrm{u}(\overline{s})) =0$. $~\hfill \square$
\end{pf}
This proposition provides sufficient conditions on the nominal trajectories and the scaling $\delta \geq V_\delta(x,z)$, such that the state/input and the corresponding geodesic satisfy the constraints. Hence, in Assumption~\ref{ass:CCM} it suffices to impose Inequalities~\eqref{eq:CCM} only on the constraint set, while most related CCM results require global assumptions~\cite{manchester2017control,singh2017robust,zhao2021tube} or additional geodesic convexity~\cite{schiller2022lyapunov}. 
The following result uses Propositions~\ref{prop:V-delta} and \ref{prop:c_j} to derive a differential equation for the tube scaling $\delta$ that accounts for the disturbance $d$ and the parametric uncertainty in $\theta$.
\begin{thm} \label{thm:tube-dyn}
	Suppose Assumption~\ref{ass:CCM} holds. Consider any initial state $x_0 \in \R^n$, nominal parameter $\overline{\theta}\in\Theta_0$, and trajectories $z_t$, $v_t$, $\delta_t$, $t\geq 0$ that satisfy
	\begin{subequations}
	\begin{align}
	  &h_j(z_t,v_t) + c_j \delta_t \leq 0, \quad j \in \I{1,r}, \label{eq:prop-constraint-tight} \\
	  &\dot{z}_t = \overline{f}(z_t,v_t,\overline{\theta}), \label{eq:prop-z-dot}\\
	  &V_\delta(x_0,z_0) \leq \delta_0, \label{eq:prop-delta_0}\\
	  &\dot{\delta}_t = -\left (\rho_\mathrm{c} - L_\D\right)\delta_t + \max_{\substack{i \in \I{1,n_{\Theta_0}} \\ j \in \I{1,n_\D}}} \left ( \sum_{k=1}^p L_{\mathrm{G},k}|[\theta^i - \overline{\theta}]_k| \delta_t \right. \nonumber \\
	  & \qquad \underbrace{\left. \vphantom{\sum_0^p}+\|G(z_t,v_t)(\theta^i - \overline{\theta}) + E(z_t)d^j\|_{M(z_t)} \right), \qquad \quad }_{:=f_\delta(\delta_t,z_t,v_t,\Theta_0,\overline{\theta})} \label{eq:f-delta-def}
	\end{align}
	for all $t \geq 0$, with $\theta^i$, $i \in \I{1,n_{\Theta_0}}$ and $d^j$, $j \in \I{1,n_\D}$ denoting the vertices of $\Theta_0$ and $\D$. Then, for any $\overline{\theta}, \theta \in \Theta_0$, $d(t) \in \D$, $t\geq 0$, the trajectory $x(0) = x_0$, $\dot{x}(t) = f_\mathrm{w}(x(t),\kappa(x(t),z_t,v_t),\theta,d(t))$, $t\geq 0$, satisfies
	\begin{align} 
	  V_\delta(x(t),z_t) \leq & \delta_t, \quad t\geq 0, \label{eq:delta-larger} \\
	  (x(t),\kappa(x(t),z_t,v_t)) \in & \Z, \quad t \geq 0. \label{eq:tube-constraint-sat}
	\end{align}
	\end{subequations}
\end{thm}
\begin{pf}
  First, note that \eqref{eq:delta-larger} holds at time $t=0$ due to Condition~\eqref{eq:prop-delta_0}. For contradiction, suppose that there exists a time $\tau \geq 0$, such that \eqref{eq:delta-larger} holds for all $t \in [0,\tau]$, but is violated for $t>\tau$, i.e., $V_\delta(x(\tau),z_\tau) = \delta_\tau$ and $\dot{V}_\delta(x(\tau),z_\tau) > \dot{\delta}_\tau$. Given that \eqref{eq:delta-larger} holds for $t =\tau$, then Proposition \ref{prop:c_j} in combination with \eqref{eq:prop-constraint-tight} implies $(\gamma^\star(s),\gamma^\mathrm{u}(s))\in \Z$, $s \in [0,1]$, with $\gamma^\star \in \Gamma(z_\tau,x(\tau)), \gamma^\mathrm{u}$ based on Equations~\eqref{eq:V-delta-def} and \eqref{eq:kappa-def}. Hence, we can invoke Inequality~\eqref{eq:Vd-contraction} in Proposition~\ref{prop:V-delta}, yielding
  \begin{align*}
    \dot{V}_\delta(x(\tau),z_\tau) \stackrel{\eqref{eq:Vd-contraction},\eqref{eq:f-delta-def}}{\leq}& f_\delta(V_\delta(x(\tau),z_\tau),z_\tau,v_\tau,\Theta_0,\overline{\theta}) \\
    =& f_\delta(\delta_\tau,z_\tau,v_\tau,\Theta_0,\overline{\theta}) \stackrel{\eqref{eq:f-delta-def}}{=} \dot{\delta}_\tau,
  \end{align*}
  which yields a contradiction. Hence, \eqref{eq:delta-larger} holds for $t\geq 0$. Furthermore, Proposition~\ref{prop:c_j} and Inequality~\eqref{eq:prop-constraint-tight} imply Inequality~\eqref{eq:tube-constraint-sat}. $\hfill \square$
\end{pf}
Theorem~\ref{thm:tube-dyn} is applicable if the nominal trajectories $z_t,v_t$ and the tube scaling $\delta_t$ are computed according to Conditions~\eqref{eq:prop-constraint-tight}--\eqref{eq:f-delta-def}. In that case, all uncertain trajectories $x(t)$, with feedback $\kappa(x(t),z_t,v_t)$ are confined within the tube $\mathbb{T}_t$ given by \eqref{eq:tube-def} (cf. Eq.~\eqref{eq:delta-larger}) and satisfy the constraints (cf. Eq.~\eqref{eq:tube-constraint-sat}). 
Thus, we obtained simpler sufficient conditions (cf. Eq.~\eqref{eq:prop-constraint-tight}--\eqref{eq:f-delta-def}) to compute a feasible solution to Problem~\eqref{eq:conceptual}. 

\subsection{Parameter update using set-membership estimation} \label{sec:set-update}
We use set-membership estimation to update the polytopic parameter set $\Theta_0$ online. This method is widely used in related RAMPC schemes~\cite{kohler2019linear,kohler2020RAMPC,lu2021robust,lorenzen2019robust,lopez2019adaptive}, since it allows for reduced conservatism. 

For the considered continuous-time setting, we assume that the state $x(t)$ and input $u(t)$ can be measured exactly, and a noisy measurement of the state derivative is available, i.e., $\dot{\tilde{x}}(t) = \dot{x}(t) + \epsilon(t)$, where $\epsilon(t) \in \D_{\epsilon}$, $t\geq 0$, with some polytope $\D_\epsilon$. Then, the non-falsified parameter set given $(x,u,\dot{\tilde{x}})$ at time t is computed as
\begin{align*}
\Delta_{t} := \{ \theta \in \R^p \big| \exists (d,\epsilon) \in \D \times \D_{\epsilon}: \dot{\tilde{x}}(t) = \dot{x}(t) + \epsilon, &\\
\dot{x}(t) = f_\mathrm{w}(x(t),u(t),\theta,d)& \}.
\end{align*}
We update the parameter set at each sampling time $t_k$, $k \in \mathbb{I}_{\geq 1}$ using $n_\mathrm{m} \in \mathbb{I}_{\geq 1}$ measurements between $t_{k-1}$ and $t_{k}$, using the following polytopic set-intersection:
\begin{equation} \label{eq:Theta-update}
\Theta_{t_{k}} = \Theta_{t_{k-1}} \cap_{j \in \I{0,n_\mathrm{m}-1}} \Delta_{t_{k-1} + j \frac{T_\mathrm{s}}{n_\mathrm{m}}}, \quad k\in\mathbb{I}_{\geq 1}.
\end{equation}
\begin{prop} \label{prop:Theta-update}
The set $\Theta_{t_{k}}$ computed according to \eqref{eq:Theta-update} is a polytope, which satisfies
    \begin{equation} \label{eq:Theta_t-prop}
        \theta \in \Theta_{t_{k}} \subseteq \Theta_{t_{k-1}} \subseteq \Theta_0, \quad k \in \mathbb{I}_{\geq 1}.
    \end{equation}
\end{prop}
\begin{pf}
Since $\Theta_{t_{k-1}}$, $\D$, and $\D_\epsilon$ are ploytopes, the sets $\Delta_{t}$, $t \geq 0$ are polyhedrons, and hence $\Theta_{t_{k}}$, $k \in \mathbb{I}_{\geq 1}$ are also polytopes. The condition $\theta \in \Theta_{t_{k}} \subseteq \Theta_{t_{k-1}}$ follows for all $k \in \mathbb{I}_{\geq 1}$ recursively, using the fact that $\theta \in \Theta_{0}$, and the definition of the non-falsified sets $\Delta_{t}$. $\hfill \qed$
\end{pf}

In contrast to existing nonlinear RAMPC schemes with set-membership estimation~\cite{lopez2019adaptive,kohler2020RAMPC}, the presented framework allows for the general update of the uncertainty set (cf. Eq.~\eqref{eq:Theta-update}) without further restrictions.
Also note that the set $\Theta_{t_k}$ is non-increasing, but may not converge to $\{\theta\}$ in general. Sufficient conditions for convergence can be found in~\cite[Cor.~8]{lu2021robust}, assuming persistence of excitation and tight characterization of the disturbance bound.

\begin{rem}
Computing the intersection of finitely many polytopes in \eqref{eq:Theta-update} is straightforward. However, the number of facets and vertices of the parameter set $\Theta_{t_k}$ might grow with the updates leading to increasing computational complexity. This issue can be addressed by considering fixed-complexity polytopes of the form $\Theta_{t_k} = \{\theta \in \R^p | H \theta \leq h_{t_k}\}$, with some fixed matrix $H \in \R^{n_\mathrm{p} \times p}$, and only update the vector $h_{t_k} \in \R^{n_p}$. This update retains the theoretical properties \eqref{eq:Theta_t-prop} and can be formulated as a linear program, see~\cite{lorenzen2019robust}.
\end{rem}

\begin{rem}
A noisy derivative $\dot{\tilde{x}}$ can be obtained using suitable filters~ \cite{Adetola2009adaptive,guay2015robust,lopez2019adaptive}.
Equivalently, the non-falsified set can be written using an integral form~\cite[Eq.~(17)]{cohen2022integralSetID}. 
Note that instead of set-membership estimation, the parameter sets $\Theta_t$ can, e.g., also be updated using Lyapunov-type arguments~\cite{Adetola2009adaptive,guay2015robust} as long as Condition~\eqref{eq:Theta_t-prop} remains valid.
\end{rem}

\subsection{Terminal set constraint} \label{sec:terminal}
In order to derive recursive feasibility and convergence guarantees of the proposed MPC scheme, we require the following standard conditions on the terminal ingredients (terminal set, terminal control law, terminal cost). In Proposition~\ref{prop:term-eq} below, we also propose a simple constructive design using a terminal equality constraint that satisfies these conditions.

\begin{assum} \label{ass:terminal}
	There exists a terminal set $\Xf \subseteq \R^n \times \R_{\geq 0} \times \Theta_0 \times 2^{\R^p}$, a terminal control law $k_{\mathrm{f}}:\R^n \times \Theta_0 \rightarrow \R^m$, and a continuous terminal cost $\ell_{\mathrm{f}}: \R^n \times \Theta_0 \rightarrow \R_{\geq 0}$, such that for any $(z,\delta,\overline{\theta},\Theta) \in \Xf$, 
	the trajectories $\dot{z}_\tau = \overline{f}(z_\tau,k_\mathrm{f}(z_0,\overline{\theta}),\overline{\theta})$, $\dot{\delta}_\tau = f_\delta(\delta_\tau,z_\tau,k_\mathrm{f}(z_0,\overline{\theta}),\Theta,\overline{\theta})$, $\tau\in[0,T_{\mathrm{s}}]$, with $\delta_0 = \delta$, $z_0 = z$ satisfy
	\begin{subequations}
            \begin{itemize}
                \item \textit{positive invariance:}
                \begin{equation}
                    (z_{T_\mathrm{s}}, \delta_{T_\mathrm{s}}, \overline{\theta}, \Theta) \in \Xf, \label{eq:ass-term-PI}
                \end{equation}
                \item \textit{constraint satisfaction:}
                \begin{equation}
                    h_j(z_\tau,k_{\mathrm{f}}(z_0,\overline{\theta})) + c_j \delta_\tau \leq 0, \quad j \in \I{1,r}, \quad \tau \in [0,T_\mathrm{s}], \label{eq:ass-term-constraint}
                \end{equation}
                \item \textit{local control Lyapunov function:}
            \end{itemize}
            \begin{equation}
                    \int_{0}^{T_\mathrm{s}} \ell(z_{\tau},k_{\mathrm{f}}(z_0,\overline{\theta}),\overline{\theta}) \mathrm{d}\tau \leq \ell_{\mathrm{f}}(z_0, \overline{\theta}) - \ell_{\mathrm{f}}(z_{T_\mathrm{s}},\overline{\theta}) \label{eq:ass-term-convergence}.
                \end{equation}
		Furthermore, for any $\hat{\delta} \in [0,\delta]$, $\hat{\Theta} \subseteq \Theta$, it holds 
	\begin{align} \label{eq:ass-term-monoton}
	    (z,\delta,\overline{\theta},\Theta) \in \Xf \Rightarrow (z,\hat{\delta},\overline{\theta},\hat{\Theta}) \in \Xf.
	\end{align}
	\end{subequations}
\end{assum}
The requirements~\eqref{eq:ass-term-PI}-\eqref{eq:ass-term-convergence} are comparable to the standard conditions in MPC (cf.~\cite{rawlings2017model}).
Furthermore, we pose the monotonicity property~\eqref{eq:ass-term-monoton}. The terminal control input is chosen to be constant in the interval $[0,T_\mathrm{s}]$, which ensures that a piece-wise constant input parametrization can be used in the MPC. 
The following proposition shows that these conditions hold using a simple terminal equality constraint w.r.t. the desired steady-state (cf.  Sec.~\ref{sec:problem-statement}). 
\begin{prop} \label{prop:term-eq}
Suppose Assumption~\ref{ass:CCM} holds. 
Then, Assumption~\ref{ass:terminal} is satisfied with $k_\mathrm{f}(z,\overline{\theta}) = v_\mathrm{ref}(\overline{\theta})$, $\ell_\mathrm{f}(z,\overline{\theta}) \equiv 0$, and the following terminal set:
\begin{subequations}
\begin{align}
    \Xf = \left \{ \vphantom{2^{\R^p}} \right. & (z,\delta,\overline{\theta},\Theta) \in \R^n \times \R_{\geq 0} \times \Theta_0 \times \left. 2^{\R^p} \right |\\
    & z = z_\mathrm{ref}(\overline{\theta}), \label{eq:term-eq-z}\\
    & \exists \overline{\delta}_\mathrm{f} \in \R_{\geq 0}:~\delta \in [0,\overline{\delta}_\mathrm{f}], \label{eq:term-eq-delta-bar} \\
    & f_\delta(\overline{\delta}_\mathrm{f},z,k_\mathrm{f}(z,\overline{\theta}),\Theta,\overline{\theta}) \leq 0, \label{eq:term-eq-delta-PI} \\
    & \left. h_j(z,k_\mathrm{f}(z,\overline{\theta})) + c_j \overline{\delta}_\mathrm{f} \leq 0,~\forall j \in \I{1,r} \vphantom{2^{\R^p}}\right \}. \label{eq:term-eq-constraint}
\end{align}    
\end{subequations}
\end{prop}
\begin{pf}
Condition \eqref{eq:term-eq-z} along with the choice of $k_\mathrm{f}$ yields $z_\tau = z_\mathrm{ref}(\overline{\theta})$ for $\tau \in [0,T_\mathrm{s}]$. Furthermore, $\delta_0\in[0,\overline{\delta}_\mathrm{f}]$ (cf.~\eqref{eq:term-eq-delta-bar}) and Inequality~\eqref{eq:term-eq-delta-PI} yields $\delta_\tau \in [0,\overline{\delta}_\mathrm{f}]$, $\tau \in [0,T_\mathrm{s}]$. Hence, positive invariance \eqref{eq:ass-term-PI} follows by considering the same $\overline{\delta}_\mathrm{f}$. Moreover, Condition~\eqref{eq:term-eq-constraint} and $\delta_\tau \in [0,\overline{\delta}_\mathrm{f}]$, $\tau\in[0,T_{\mathrm{s}}]$  yield Inequality \eqref{eq:ass-term-constraint}. 
Condition~\eqref{eq:ass-term-convergence} follows with $\ell(z_\tau,k_\mathrm{f}(z_0,\overline{\theta}),\overline{\theta}) = 0$ using~\eqref{eq:stage-cost-def}. The monotonicity property~\eqref{eq:ass-term-monoton} follows using Condition~\eqref{eq:term-eq-delta-bar} and the definition of $f_\delta$ \eqref{eq:f-delta-def} yielding $f_\delta(\overline{\delta}_\mathrm{f},z,k_\mathrm{f}(z,\overline{\theta}),\hat{\Theta},\overline{\theta}) \leq f_\delta(\overline{\delta}_\mathrm{f},z,k_\mathrm{f}(z,\overline{\theta}),\Theta,\overline{\theta}) \leq 0$, $\hat{\Theta}\subseteq\Theta$. $\hfill \qed$
\end{pf}
Invariance of the terminal set w.r.t. the nominal dynamics $\overline{f}$ is guaranteed by requiring the terminal state to coincide with the steady-state of the nominal dynamics $z_\mathrm{ref}$~\eqref{eq:term-eq-z} and using the corresponding input $v_\mathrm{ref}$ as the terminal control law.  
Note that the terminal set constraint includes not only the nominal state $z$, but also the tube scaling $\delta$, the nominal parameter $\overline{\theta}$, and the parameter set $\Theta$, since $\Theta$ is updated online and $\delta$, $\overline{\theta}$ are optimization variables in the proposed RAMPC framework (Sec.~\ref{sec:framework}). Positive invariance (cf.~\eqref{eq:term-eq-delta-PI}) and constraint satisfaction (cf.~\eqref{eq:term-eq-constraint}) are shown by additionally computing a constant $\overline{\delta}_\mathrm{f}\geq 0$ that upper bounds the tube scaling $\delta$ in the terminal set. For implementation, the constant $\overline{\delta}_\mathrm{f}\geq 0$ needs to be included as a decision variable in the MPC optimization problem introduced below.

\subsection{RAMPC framework and theoretical analysis} \label{sec:framework} 
In the following, we present the overall RAMPC algorithm (Alg.~\ref{alg:offline}--\ref{alg:online}), including a corresponding theoretical analysis of the closed-loop properties (Thm.~\ref{thm:main}). The RAMPC scheme approximates the solution to the infinite-horizon Problem~\eqref{eq:conceptual} by a finite prediction horizon $T_\mathrm{f}=T_\mathrm{s} N$, $N\in\mathbb{I}_{\geq 1}$ using a receding horizon implementation with a sampling period of $T_\mathrm{s}$. In particular, at time $t$, given the measured state $x(t)$ and the updated parameter set $\Theta_t$, the proposed RAMPC scheme is characterized by the following optimization problem:
\begin{subequations} \label{eq:MPC}
	\begin{align}	
	V_{T_\mathrm{f}}^\star(x(t),\Theta_t) =& \min_{\substack{v_{\cdot|t},z_{\cdot|t} \\ \delta_{\cdot|t}, \overline{\theta}_t}} \int_{0}^{T_\mathrm{f}} \ell(z_{\tau|t},v_{\tau|t},\overline{\theta}_t) \mathrm{d}\tau + \ell_{\mathrm{f}}(z_{T_\mathrm{f}|t},\overline{\theta}_t) \nonumber \\
	\text{s.t.} \quad & \dot{z}_{\tau|t}=\overline{f}(z_{\tau|t},v_{\tau|t},\overline{\theta}_t), \label{eq:MPC-dynamics}\\
	&\dot{\delta}_{\tau|t}=f_\delta(\delta_{\tau|t},z_{\tau|t},v_{\tau|t},\Theta_t,\overline{\theta}_t), \label{eq:MPC-delta}\\
	& \overline{\theta}_t \in \Theta_0, \label{eq:MPC-theta-in-Theta}\\
	&h_j(z_{\tau|t},v_{\tau|t}) + c_j \delta_{\tau|t} \leq 0, \label{eq:MPC-constraints}\\
	&V_\delta(x(t),z_{0|t}) \leq \delta_{0|t}, \label{eq:MPC-delta0} \\
	&(z_{T_\mathrm{f}|t},\delta_{T_\mathrm{f}|t},\overline{\theta}_t,\Theta_t) \in \Xf, \label{eq:MPC-terminal}\\
	&\tau \in [0,T_\mathrm{f}],\quad j \in \I{1,r}. \nonumber
	\end{align}
\end{subequations}
The decision variables are the nominal state and input trajectories $z_{\cdot|t}$, $v_{\cdot|t}$, the tube scaling $\delta_{\cdot|t}$, and the nominal parameter $\overline{\theta}_t$. The tube scaling $\delta_{\tau|t}$ is propagated according to Theorem~\ref{thm:tube-dyn} (cf.~\eqref{eq:MPC-delta}, \eqref{eq:MPC-delta0}) and the nominal trajectory $z_{\tau|t}$ satisfies the nominal dynamics $\overline{f}$ \eqref{eq:MPC-dynamics}. The constraint tightening~\eqref{eq:MPC-constraints} ensures that the true state and input trajectory satisfy the constraints~\eqref{eq:constr}  (cf. Prop.~\ref{prop:c_j} and Thm.~\ref{thm:tube-dyn}). Condition~\eqref{eq:MPC-theta-in-Theta} allows to additionally optimize over the nominal parameter that influences the dynamics in~\eqref{eq:MPC-dynamics}--\eqref{eq:MPC-delta}. Furthermore, Condition~\eqref{eq:MPC-terminal} captures the terminal constraint. 
For simplicity, we parameterize the nominal input $v_{\tau|t}$ as piece-wise constant in each interval $\tau \in [jT_\mathrm{s},(j+1)T_\mathrm{s})$, $j \in \I{0,N-1}$. 

Appropriate conditions ensuring that a minimizer to Problem~\eqref{eq:MPC} exists can be found in \cite[Prop. 2]{fontes2001general}. We assume that a corresponding unique minimizer exists, which is denoted by the superscript $\star$. We solve the problem at each sampling time $t_k$, $k \in \mathbb{I}_{\geq 0}$, and the resulting closed-loop system is given by
\begin{align} \label{eq:closed-loop}
\begin{split}
  u(t) =& \kappa(x(t),z_{t-t_k|t_k}^*,v_{t-t_k|t_k}^*), \quad t \in [t_k,t_{k+1}), \\
  \dot{x}(t) =& f_{\mathrm{w}}(x(t),u(t),\theta,d(t)).
\end{split}
\end{align}
The offline design and the online operation are captured in the algorithms below.

\begin{algorithm}[H]
\caption{Offline design}\label{alg:offline}
\begin{algorithmic}[1]
\Statex Given model with uncertainty characterization $\D$, $\Theta_0$, $\mathbb{D}_{\epsilon}$, constraints~\eqref{eq:constr}, desired setpoint $z_\mathrm{ref}(\overline{\theta})$,  $v_\mathrm{ref}(\overline{\theta})$ and stage cost weighting $Q,R \succeq 0$~\eqref{eq:stage-cost-def}.
\State Compute $M(x)$, $K(x)$ and $\rho_\mathrm{c}$ (Asm.~\ref{ass:CCM}, Rk.~\ref{rem:CCM-computation}).
\State Compute constants $L_\D$~\eqref{eq:app-L-D}, $L_{\mathrm{G},k}$~\eqref{eq:app-L-G}, $c_j$~\eqref{eq:c_j-def}.
\State Set sampling period $T_\mathrm{s}$ and prediction horizon $T_\mathrm{f}$.
\State Design terminal ingredients (Asm.~\ref{ass:terminal}, Prop.~\ref{prop:term-eq}).
\end{algorithmic}
\end{algorithm}

\begin{algorithm}[H]
\caption{Online operation}\label{alg:online}
\begin{algorithmic}
\For{each sampling time $t_k = k T_\mathrm{s}$, $k \in \mathbb{I}_{\geq 0}$}
\State Update $\Theta_{t_k}$ \eqref{eq:Theta-update} using measurements ($x,\dot{\tilde{x}},u$).
\State Solve Problem \eqref{eq:MPC}.
\State Apply feedback $\kappa$ over period $t \in [t_k,t_{k+1})$~\eqref{eq:closed-loop}.
\EndFor
\end{algorithmic}
\end{algorithm}

The following theorem summarizes the theoretical properties of the proposed RAMPC algorithm.
\begin{thm} \label{thm:main}
	Let Assumptions~\ref{ass:CCM} and \ref{ass:terminal} hold. Suppose that Problem \eqref{eq:MPC} is feasible at time $t=0$ with initial state $x_0$ and parameter set $\Theta_{0}$. Then, Problem \eqref{eq:MPC} is feasible for all sampling times $t_k$, $k \in \mathbb{I}_{\geq 0}$, and the closed-loop system~\eqref{eq:closed-loop} resulting from Algorithm~\ref{alg:online} satisfies the constraints \eqref{eq:constr} for all $t \geq 0$. Furthermore, the resulting nominal trajectories converge to the reference state and input, i.e., $\lim_{k \to \infty}\left\|(z_{\tau|t_k}^\star,v_{\tau|t_k}^\star) - (z_\mathrm{ref}(\overline{\theta}_{t_k}^\star),v_\mathrm{ref}(\overline{\theta}_{t_k}^\star)) \right\| = 0$, $\tau \in [0,T_\mathrm{s})$.
\end{thm}

\begin{pf}
\textbf{Part I:} Assume that Problem~\eqref{eq:MPC} is feasible at some time $t_k$, $k\in \mathbb{I}_{\geq 0}$. For simplicity, we define $v_{\tau|t_k}^\star := k_{\mathrm{f}}(z_{T|t_k}^\star,\overline{\theta}_{t_k}^\star)$ and $z_{\tau|t_k}^\star$, $\delta_{\tau|t_k}^\star$ according to \eqref{eq:MPC-dynamics} and \eqref{eq:MPC-delta} for $\tau \in [T_\mathrm{f},T_\mathrm{f}+T_\mathrm{s}]$. At time $t_{k+1}$, consider the candidate solution 
\begin{align*}
\overline{\theta}_{t_{k+1}} &= \overline{\theta}_{t_k}^\star \in \Theta_0, \quad v_{\tau|t_{k+1}} = v_{\tau + T_\mathrm{s}|t_k}^\star, \quad \tau \in [0,T_\mathrm{f}], \\
\quad z_{0|t_{k+1}} & = z_{T_\mathrm{s}|t_k}^\star, \quad \delta_{0|t_{k+1}} = V_\delta(x(t_{k+1}),z_{T_\mathrm{s}|t_k}^\star),
\end{align*}
with trajectories $z_{\tau|t_{k+1}}$, $\delta_{\tau|t_{k+1}}$, $\tau \in (0,T_\mathrm{f}]$ according to the constraints \eqref{eq:MPC-dynamics}, \eqref{eq:MPC-delta}, which implies $z_{\tau|t_{k+1}} = z_{\tau+T_\mathrm{s}|t_k}^\star$, $\tau \in~[0,T_\mathrm{f}]$. In the following, we show that the tube scaling satisfies the following nestedness property:
\begin{equation} \label{eq:T-cont-proof-delta-nested}
\delta_{\tau|t_{k+1}} \leq \delta_{\tau+T_\mathrm{s}|t_k}^\star, \quad \tau \in [0,T_\mathrm{f}].
\end{equation}
First, note that Condition~\eqref{eq:delta-larger} in Theorem~\ref{thm:tube-dyn} also holds for any $\Theta \subseteq \Theta_0$ with $\dot{x} = f_\mathrm{w}(x,u,\theta,d)$, $\dot{\delta}=f_\delta(\delta,z,v,\Theta,\overline{\theta})$, $\overline{\theta}\in\Theta_0$, $\theta \in \Theta$, $d \in \D$ yielding
\begin{equation*}
V_\delta(x(t_k+\tau),z_{\tau|t_k}^\star) \leq \delta_{\tau|t_k}^\star, \quad \tau \in [0,T_\mathrm{s}].
\end{equation*}
Hence, the initial value of the tube scaling satisfies
\begin{equation*}
\delta_{0|t_{k+1}} = V_\delta(x(t_{k+1}),z_{T_\mathrm{s}|t_k}^\star) \leq \delta_{T_\mathrm{s}|t_k}^\star.
\end{equation*}
For contradiction, suppose that there exists a time $\hat{\tau} \in [0,T_\mathrm{f})$, such that \eqref{eq:T-cont-proof-delta-nested} holds for all $\tau \in [0,\hat{\tau}]$, but is violated for $\tau > \hat{\tau}$, i.e., $\delta_{\hat{\tau}|t_{k+1}} = \delta_{\hat{\tau}+T_\mathrm{s}|t_k}^\star$ and $\dot{\delta}_{\hat{\tau}|t_{k+1}} > \dot{\delta}_{\hat{\tau}+T_\mathrm{s}|t_k}^\star$. The dynamics of the tube scaling~\eqref{eq:f-delta-def} in combination with $\Theta_{t_{k+1}} \subseteq \Theta_{t_k}$ yield
\begin{align*}
  \dot{\delta}_{\hat{\tau}|t_{k+1}}=& f_\delta(\delta_{\hat{\tau}|t_{k+1}},z_{\hat{\tau}+T_\mathrm{s}|t_k}^\star,v_{\hat{\tau}+T_\mathrm{s}|t_k}^\star,\Theta_{t_{k+1}},\overline{\theta}_{t_k}^\star) \\
  \leq & f_\delta(\delta_{\hat{\tau}+T_\mathrm{s}|t_k}^\star,z_{\hat{\tau}+T_\mathrm{s}|t_k}^\star,v_{\hat{\tau}+T_\mathrm{s}|t_k}^\star,\Theta_{t_k},\overline{\theta}_{t_k}^\star) \\
 = & \dot{\delta}_{\hat{\tau}+T_\mathrm{s}|t_k}^\star,
\end{align*}
which is a contradiction, and hence~\eqref{eq:T-cont-proof-delta-nested} holds. \\
The candidate solution satisfies Condition~\eqref{eq:MPC-terminal} due to the invariance~\eqref{eq:ass-term-PI} and monotonicity property~\eqref{eq:ass-term-monoton} of $\Xf$, using Inequality~\eqref{eq:T-cont-proof-delta-nested} and $\Theta_{t_{k+1}}\subseteq\Theta_{t_k}$. Inequality~\eqref{eq:MPC-constraints} for $\tau \in [T_\mathrm{f}-T_\mathrm{s},T_\mathrm{f}]$ follows analogously using Condition~\eqref{eq:ass-term-constraint}. For $\tau \in [0,T_\mathrm{f} - T_\mathrm{s}]$, Condition~\eqref{eq:MPC-constraints} holds with:
\begin{align*}
& h_j(z_{\tau|t_{k+1}}, v_{\tau|t_{k+1}}) + c_j \delta_{\tau|t_{k+1}} \\ 
& \stackrel{\eqref{eq:T-cont-proof-delta-nested}}{\leq} h_j(z_{\tau+T_\mathrm{s}|t_k}^\star, v_{\tau+T_\mathrm{s}|t_k}^\star) + c_j \delta_{\tau+T_\mathrm{s}|t_k}^\star \stackrel{\eqref{eq:MPC-constraints}}{\leq} 0,
\end{align*}
where we used feasibility of Problem~\eqref{eq:MPC} at time $t_k$ and the fact that $c_j \geq 0$. \\
\textbf{Part II:} Constraint satisfaction~\eqref{eq:constr} for the closed-loop system follows from Theorem~\ref{thm:tube-dyn} using the tightened constraints~\eqref{eq:MPC-constraints} for $\tau\in[0,T_\mathrm{s}]$ and the tube dynamics~\eqref{eq:MPC-delta0}, \eqref{eq:MPC-delta}. \\
\textbf{Part III:} 
Convergence is established using standard inequalities for the optimal cost $V_{T_\mathrm{f}}^\star$. In particular, the feasible candidate solution provides an upper bound to the optimal cost, i.e.,
\begin{equation*}
\begin{split}
&V_{T_\mathrm{f}}^\star(x(t_{k+1}),\Theta_{t_{k+1}})\\
\leq & \int_{0}^{T_\mathrm{f}} \ell(z_{\tau+T_\mathrm{s}|t_k}^\star, v_{\tau+T_\mathrm{s}|t_k}^\star,\overline{\theta}_{t_k}^\star) + \ell_{\mathrm{f}}(z_{T_\mathrm{f}+T_\mathrm{s}|t_k}^\star,\overline{\theta}_{t_k}^\star) \mathrm{d}\tau\\
= & V_{T_\mathrm{f}}^\star(x(t_k),\Theta_{t_k}) - \int_0^{T_\mathrm{s}} \ell(z_{\tau|t_k}^\star,v_{\tau|t_k}^\star,\overline{\theta}_{t_k}^\star)\mathrm{d}\tau \\
&- \ell_{\mathrm{f}}(z_{T_\mathrm{f}|t_k}^\star,\overline{\theta}_{t_k}^\star) + \int_{T_\mathrm{f}}^{T_\mathrm{f}+T_\mathrm{s}} \ell(z_{\tau|t_k}^\star,v_{\tau|t_k}^\star,\overline{\theta}_{t_k}^\star)\mathrm{d}\tau \\
&+ \ell_{\mathrm{f}}(z_{T_\mathrm{f}+T_\mathrm{s}|t_k}^\star,\overline{\theta}_{t_k}^\star) \\
\stackrel{\eqref{eq:ass-term-convergence}}{\leq} & V_{T_\mathrm{f}}^\star(x(t_k),\Theta_{t_k}) - \int_0^{T_\mathrm{s}} \ell(z_{\tau|t_k}^\star,v_{\tau|t_k}^\star,\overline{\theta}_{t_k}^\star)\mathrm{d}\tau.
\end{split}
\end{equation*}
Applying the previous inequality recursively yields
\begin{align*}
&V_{T_\mathrm{f}}^\star(x_0,\Theta_0) - \limsup_{k \rightarrow \infty}V_{T_\mathrm{f}}^\star(x(t_k),\Theta_{t_k}) \\
\geq &\sum_{k=0}^\infty \int_0^{T_\mathrm{s}} \ell(z_{\tau|t_k}^\star,v_{\tau|t_k}^\star,\overline{\theta}_{t_k}^\star)\mathrm{d}\tau.
\end{align*}
Compact constraints in combination with a continuous cost imply a uniform bound on $V_{T_\mathrm{f}}^\star$, and hence the right hand side of the above inequality is bounded, yielding 
\begin{align*}
\lim_{k\to \infty} \int_0^{T_\mathrm{s}} \ell(z_{\tau|t_k}^\star,v_{\tau|t_k}^\star,\overline{\theta}_{t_k}^\star)\mathrm{d}\tau = 0. 
\end{align*}
Note that $z_{\tau|t_k}^\star$ is uniformly continuous in $\tau$, given $\overline{f}$ Lipschitz and $(z_{\tau|t_k}^\star,v_{\tau|t_k}^\star)$, $\overline{\theta}^\star_{t_k}$ subject to compact constraints~\eqref{eq:MPC-constraints}, \eqref{eq:MPC-theta-in-Theta}. Furthermore, $v_{\tau|t_k}^\star$ and $\overline{\theta}^\star_{t_k}$ are constant on $\tau \in [0,T_\mathrm{s})$, thus $\ell(z_{\tau|t_k}^\star,v_{\tau|t_k}^\star,\overline{\theta}_{t_k}^\star)$ is uniformly continuous on $\tau \in [0,T_\mathrm{s})$. Since $\ell$ is also positive definite, Barbalat's Lemma \cite{khalil2002nonlinear} can be invoked yielding:
\begin{align*}
\lim_{k \to \infty} \left \|(z_{\tau|t_k}^\star, v_{\tau|t_k}^\star) - (z_\mathrm{ref}(\overline{\theta}^\star_{t_k}), v_\mathrm{ref}(\overline{\theta}^\star_{t_k})) \right\| = 0. \qed
\end{align*}
\end{pf}
Theorem~\ref{thm:main} provides all the desired properties of a robust MPC scheme, while incorporating online updates of the parameter set $\Theta_t$ to reduce conservatism. 

\begin{rem} \label{rem:implementation}
The dynamics~\eqref{eq:MPC-delta} involve the maximization operator from $f_\delta$~\eqref{eq:f-delta-def}, which may complicate integration. This issue is circumvented by introducing an auxiliary variable $\overline{w}_{\tau|t} \geq 0$ and using the following constraints:
\begin{align}
\dot{\delta}_{\tau|t}= &-(\rho_\mathrm{c} - L_\D)\delta_{\tau|t}+\overline{w}_{\tau|t}, \label{eq:w_bar} \nonumber\\
\overline{w}_{\tau|t} \geq& \sum_{k=1}^p L_{\mathrm{G},k}|[\theta^i - \overline{\theta}_t]_k| \delta_{\tau|t} +\|G(z_{\tau|t},v_{\tau|t})(\theta^i - \overline{\theta}_t) \nonumber \\
& + E(z_{\tau|t})d^j\|_{M(z_{\tau|t})}, \quad i \in \I{1,n_{\Theta_t}}, \quad j \in \I{1,n_\D}. 
\end{align}
Hence, we can equivalently write the dynamics~\eqref{eq:MPC-dynamics}, \eqref{eq:MPC-delta} and the constraints~\eqref{eq:MPC-constraints} using smooth continuous-time dynamics with state $(z,\delta) \in \R^{n+1}$, input $(v,\overline{w})\in\R^{m+1}$, and continuous-time constraints~\eqref{eq:MPC-constraints}, \eqref{eq:w_bar}, analogous to standard continuous-time MPC schemes~\cite[Fig. 12.10]{grune2017nonlinear}. Correspondingly, we use standard methods to integrate the dynamics. Furthermore, we relax the continuous-time constraint satisfaction by only enforcing constraints on some discrete points~\cite[Sec. 8]{rawlings2017model}. 
However, one could also use additional modifications to rigorously guarantee continuous-time constraint satisfaction (cf. \cite[Thm.~3]{magni2004model}, \cite{fontes2018guaranteed}). \\
Note that constraint~\eqref{eq:MPC-delta0} on the initial state can be equivalently written as $\int_0^1\|\gamma_{\mathrm{s}}(s)\|_{M(\gamma(s))}\mathrm{d}s \leq \delta_{0|t}$ (cf. Eq.~\eqref{eq:V-delta-def}) by using the curve $\gamma \in \Gamma(z_{0|t},x(t))$ as an additional optimization variable. The integral is approximated with numerical quadrature, and the curve $\gamma$ is finitely parametrized using a pseudospectral method (cf., Remark~\ref{rem:CCM-computation} and \cite{leung2017nonlinear,zhao2021tube}). Thus, Problem~\eqref{eq:MPC} is approximated with a finite dimensional nonlinear program.
\end{rem}

\subsection{Discussion} \label{sec:discussion} 
In the following, we discuss the results and benefits of the proposed RAMPC framework in comparison to existing robust and robust adaptive MPC schemes.

\textit{Stability properties}: 
Suppose for simplicity that $z_\mathrm{ref}(\overline{\theta})$, $v_\mathrm{ref}(\overline{\theta})$ is unique. Then, the convergence result in Theorem~\ref{thm:main} yields that the uncertain closed-loop trajectory $x(t)$ converges to an RPI set corresponding to the stabilizing feedback $u(t)=\kappa\left(x(t),z_\mathrm{ref}(\overline{\theta}),v_\mathrm{ref}(\overline{\theta})\right)$, analogous to standard tube MPC results~\cite{mayne2005robust,bayer2013discrete}. To improve performance, the cost function can also be based on a state trajectory predicted according to the true measured state $x(t)$ and some other parameter estimate $\hat{\theta}_{t_k}$. Specifically, by using least mean square estimate, finite-gain $\mathcal{L}_2$-stability w.r.t. disturbance $d$ is shown in \cite{kohler2020RAMPC,kohler2019linear,lorenzen2019robust}. 

\textit{Tube propagation}: 
The tube propagation derived in Theorem~\ref{thm:tube-dyn} is similar to the RMPC formulations presented in~\cite[App. B]{Koehler2020Robust} and \cite{kohler2020RAMPC}, which use the same tube parametrization~\eqref{eq:tube-def} with $V_\delta$, $z$, $\delta$, compare also~\cite{rakovic2022homothetic}. 
However, in~\cite{Koehler2020Robust,kohler2020RAMPC} explicit design formulas for the tube propagation were only established for the special case of ellipsoidal sets (cf.~\cite[Sec.~3.5]{kohler2020RAMPC}) and \cite{rakovic2022homothetic} only studies the special case of a shift invariant parametrization.\footnote{The controller $\kappa$, the incremental Lyapunov function $V_\delta$, the tube $\mathbb{T}$, and the bound on the model mismatch may only depend on the error $x-z$, which excludes general CCMs.}
Furthermore, the simple tube propagation in~\cite{Adetola2009adaptive,guay2015robust,pin2009robust} based on Lipschitz continuity is a further special case of the proposed framework with $M(z)=I$ and $K(z)=0$, compare also~\cite[Rk. 4]{Koehler2020Robust} and \cite[Sec.~3.6]{kohler2020RAMPC} for a detailed discussion. 
In contrast to these existing RMPC approaches, we utilized general CCMs to provide simple explicit formulas for the tube dynamics and constraint tightening using the gradient theorem, compare~\eqref{eq:app-L-G}, \eqref{eq:app-L-D}, \eqref{eq:c_j-def}. \\
By using the upper bound $M(z)\preceq\overline{M}$ in Inequality~\eqref{eq:w_bar}, we can also obtain a computationally cheaper, yet more conservative result.
Moreover, in the robust case ($\overline{\theta}$, $\Theta$ constant), we can compute a constant $L \leq L_\D + \max_{i \in \I{1,n_{\Theta_0}}} \sum_{k=1}^{p} L_{\mathrm{G},k} \left | [\theta^i - \overline{\theta}]_k \right|$, which is also less conservative compared to the formulation in~\cite{Koehler2020Robust}.

\textit{Homothetic vs. rigid tube}: 
Many robust MPC formulations use a fixed (rigid) tube scaling $\delta_{\tau|t}=\overline{\delta}>0$ for all $\tau\geq0$ instead of implementing a tube propagation of the form Eq.~\eqref{eq:f-delta-def}, compare~\cite{singh2019robust,singh2017robust,wabersich2021nonlinear,zhao2021tube,bayer2013discrete,yu2013tube}. 
This special case can be recovered by computing a constant $\overline{\delta}>0$, such that $\dot{\delta}=f_\delta(\overline{\delta},z,v,\overline{\theta},\Theta)\leq 0$ for all $(z,v)\in\mathbb{Z}$. 
Similarly, an arbitrary constant $\delta_{\tau|t}=\overline{\delta}_{\max}>0$, $\tau\geq 0$, can be fixed offline by adding the constraint $\overline{w}_{\tau|t}\leq \overline{w}_{\max}:=(\rho_c-L_{\mathbb{D}})\overline{\delta}_{\max}$ to upper bound the right hand side of Inequality~\eqref{eq:w_bar} to the optimization problem. 
This restricts the nominal state and input to regions with small uncertainty, compare~\cite{wabersich2021nonlinear}.
On the other hand, the proposed homothetic tube MPC overcomes this conservatism by systematically changing the scaling $\delta$ along the horizon, compare also the numerical example in Section~\ref{sec:example}.

\textit{Initial state constraint}: 
The proposed homothetic tube MPC framework allows for an optimization over the initial state $z_{0|t}$. However, this requires an evaluation of $V_\delta$ in Condition~\eqref{eq:MPC-delta0}, which increases the computational complexity, compare Remark~\ref{rem:implementation}. 
This can be avoided by considering the more restrictive initial state constraint $z_{0|t}=x(t)$, $\delta_{0|t}=0$ (cf. Eq.~\eqref{eq:MPC-delta0}). Such an initial state constraint is also suggested in the homothetic tube MPC schemes in~\cite{Koehler2020Robust,kohler2020RAMPC}, where recursive feasibility is demonstrated using a candidate input $v$ that tracks the previous optimal solution $z^\star$.\footnote{Accounting for the piece-wise constant input parametrization requires additional modifications (cf., e.g.,~\cite[Rk.~14]{Koehler2020Robust}).} 
We note that this approach necessitates a stronger (robust) positive invariance condition on the terminal set, which excludes the simple terminal equality constraints (cf. Prop.~\ref{prop:term-eq}).

\textit{Adaptive MPC}: 
In the following, we discuss the difference in how we utilize set-membership estimation compared to existing adaptive MPC frameworks~\cite{kohler2020RAMPC,lopez2019adaptive,Adetola2009adaptive,guay2015robust}.
One main novelty in our work is the optimization over the nominal parameter $\overline{\theta}_t$, yielding a number of benefits. 
Intuitively, one might be inclined to simply fix $\overline{\theta}_t$ at the center of $\Theta_t$, as done in \cite{kohler2020RAMPC,Adetola2009adaptive,guay2015robust,lopez2019adaptive,kohler2019linear,didier2021robust}, since this results in small tube scaling $\delta_t$ (cf. Eq.~\eqref{eq:f-delta-def}).
However, this complicates the recursive feasibility analysis, limiting the applicability to restrictive parameter sets, such as hypercubes~\cite{kohler2020RAMPC,kohler2019linear,didier2021robust,lopez2019adaptive} or balls~\cite{guay2015robust,Adetola2009adaptive}. On the other hand, optimizing over $\overline{\theta}_t$ enables a simple recursive feasibility proof by shifting the previous optimal solution. 
As a result, the proposed RAMPC scheme is applicable to general CCMs (Asm.~\ref{ass:CCM}), and requires no restrictions on the parametrization in the set-membership estimation (Sec.~\ref{sec:set-update}). \\
Finally, in contrast to existing adaptive methods (cf.~\cite[App. B]{kohler2019linear}, \cite[Chap. II.D]{didier2021robust}), tracking  parameter dependent set-points $(z_{\mathrm{ref}}(\overline{\theta}),v_{\mathrm{ref}}(\overline{\theta}))$ is straightforward. 
Such a set-point can be implicitly specified using:
\begin{equation*}
\begin{split}
    (z_\mathrm{ref}(\overline{\theta}),v_\mathrm{ref}(\overline{\theta})) \in& \arg \min_{\substack{z\in \R^n \\ v \in \R^m}} \|C z + D v - y^\mathrm{d}\|^2 \\
    \textnormal{s.t.} \quad & \overline{f}(z,v,\overline{\theta}) = 0, \quad (z,v) \in \Z,
\end{split}
\end{equation*}
where $y^\mathrm{d}$ is a reference for a system output $y=Cx+Du$. This optimization problem can also be included in the MPC formulation, similar to~\cite{limon2018nonlinear}.

\section{Numerical example} \label{sec:example}
The following example demonstrates the benefits of the proposed homothetic tube MPC framework and the online model updates. We consider the following planar quadrotor from \cite{singh2017robust, zhao2021tube}:
\begin{equation*}
{\scriptsize
\begin{bmatrix}
\dot{p}_1 \\ \dot{p}_2 \\ \dot{\phi} \\ \dot{v}_1 \\ \dot{v}_2 \\ \ddot{\phi}
\end{bmatrix}
=
\begin{bmatrix}
v_1 \cos(\phi) - v_2 \sin(\phi) \\
v_1 \sin (\phi) + v_2 \cos(\phi) \\
\dot{\phi} \\
v_2 \dot{\phi} - g \sin(\phi) \\
-v_1 \dot{\phi} - g \cos(\phi) \\
0
\end{bmatrix}
+
\begin{bmatrix}
0 & 0 \\
0 & 0 \\
0 & 0 \\
0 & 0 \\
\frac{1}{m} & \frac{1}{m} \\
\frac{l}{J} & \frac{-l}{J} \\
\end{bmatrix}
u +  
\begin{bmatrix}
0\\0\\0\\\cos(\phi)\\ -\sin(\phi) \\ 0
\end{bmatrix}
d,}
\end{equation*}
where $p_1, p_2$ are horizontal and vertical positions, $v_1, v_2$ are velocities in the body frame, and $\phi,\dot{\phi}$ denote the angle/angular velocity. The control input $u=[u_1,u_2]^\top$ is the thrust force produced by the propellers and $d$ is a wind disturbance. Furthermore, $g$, $m$, and $l$ denote the gravitational acceleration, mass, and the distance between each of the propellers and the vehicle center, respectively. Compared to the setting in~\cite{singh2017robust,zhao2021tube}, we additionally consider the mass as an uncertain parameter $\theta=\frac{1}{m}$ with initial estimate $\overline{\theta}_0 = 2.058$, uncertainty set $\Theta_0 = [0.99\overline{\theta}_0, 1.01\overline{\theta}_0]$, and true parameter $\theta=1.01\overline{\theta}_0$. The disturbance satisfies $d(t)\in \D := [-0.1,0.1]$. The state and input constraint set~\eqref{eq:Z_set} are given by:
$|\phi| \leq \pi/3$, $|\dot{\phi}| \leq \pi$, $|v_1| \leq 2$, $|v_2| \leq 1$, $u_i\in[-1,3.5]$, $i\in\{1,2\}$. 
The goal is to reach position $p=0$, which corresponds to  $z_\mathrm{ref}(\overline{\theta}) = 0$,  $v_\mathrm{ref}(\overline{\theta}) = \frac{g}{2\overline{\theta}} [1, 1]^\top$.  
We have placed circular obstacles between the initial position and the goal in order to provide a challenging control problem, compare Figures~\ref{fig:A-R} and \ref{fig:R-CB}.
Note that planning robust trajectories in this cluttered environment requires small tubes, which is only attainable with a relatively small uncertainty sets $\mathbb{D},\Theta_0$. 
Handling significantly larger uncertainty with the proposed framework would lead to feasibility issues.
Nonetheless, the proposed approach is significantly less conservative than existing methods~\cite{singh2017robust,zhao2021tube} as shown in the subsection \textit{Comparison to rigid tube MPC} below.

\begin{rem}
Applying the presented RAMPC scheme for any system is straightforward in case a valid CCM is available.
This includes any feedback linearizable system (cf.~\cite{manchester2017control}) or any system that is quadratically stabilizable, i.e., admits a constant CCM.
In particular, CCMs have been derived for a large class of nonlinear systems in the literature, including e.g., 3D quadrotors~\cite{zhao2021tube}, cars~\cite{chou2023safe,kohler2020reference}, manipulators~\cite{chou2023safe}, and chemical reactors~\cite{kohler2020reference}.
\end{rem}

\subsubsection*{Implementation details}
The CCM (Asm.~\ref{ass:CCM}) is computed using the convex re-parametrization $Y$, $W$ (cf. Remark~\ref{rem:CCM-computation}) by adjusting the SOS code from~\cite{zhao2021tube}.  
We optimize for a CCM that results in a smaller tube by adding the following LMI
\begin{equation} \label{eq:extra-LMI}
    \begin{bmatrix}
    W & G\tilde{\theta} + Ed \\ (G\tilde{\theta} + Ed)^\top & \tilde{w}^2_{\max} 
    \end{bmatrix}
    \succeq 0, \quad \forall (d,\tilde{\theta}+\overline{\theta}_0)\in\mathbb{D}\times\Theta_0
\end{equation}
and minimizing $\tilde{w}_{\max}$.\footnote{Applying the Schur complement to~\eqref{eq:extra-LMI} yields $\tilde{w}_{\max}^2 \geq \|G\tilde{\theta} + Ed\|_M^2$ which provides an upper bound to the last term in $f_\delta$~\eqref{eq:f-delta-def}.}
We note that simpler RMPC schemes based on ellipsoidal/polytopic RPI sets  (cf., e.g.,~\cite{yu2013tube,wabersich2021nonlinear,rakovic2022homothetic}, corresponding to constant matrices $Y,W$) or feedback linearization (cf.~\cite{lopez2019adaptive}) are not feasible for this nonlinear problem.
In particular, if the linearized model around an equilibrium would be used for a standard linear robust MPC design~\cite{chisci2001systems}, then the constraint tightening on the horizontal position would be over 100 meters at the end of the prediction horizon, i.e., multiple factors too large to apply to the cluttered environment in Figure~\ref{fig:A-R}.
This shows that the problem is in fact highly nonlinear and requires nonlinear robust MPC techniques to achieve good performance.
\\
The dynamics~\eqref{eq:MPC-dynamics}, \eqref{eq:MPC-delta} as well as the true dynamics~\eqref{eq:system} are discretized using explicit Runge-Kutta 4 discretization with the sampling interval $T_\mathrm{s} = 150$ ms, and we only enforce the state and input constraints at the sampling times $t_k$ (cf. Remark~\ref{rem:implementation}). We choose a prediction horizon of $T_{\mathrm{f}}=N \cdot T_{\mathrm{s}}=3.75$ s based on $N=25$ steps. The disturbance $d(t)$ and the measurement noise $\epsilon(t)$ are sampled uniformly from the sets $\D$ and $\D_\epsilon = \D$. The geodesic in~\eqref{eq:MPC-delta0} is approximated using Chebyshev interpolating polynomials up to degree 2 (cf. Remark~\ref{rem:implementation}) with the code from~\cite{zhao2021tube}. \\
All computations were carried out using Matlab, on a Dell Inspiron 15-3567 laptop with Intel i7-7500U CPU and 8 GB RAM running Windows 10. The MPC problems are solved with IPOPT \cite{wachter2006implementation} formulated in CasADi \cite{andersson2019casadi}.\footnote{To avoid local minima, we solve the resulting NLP with initial guesses based on different paths around the obstacles.} The corresponding code is available online.\footnote{gitlab.ethz.ch/ics/RAMPC-CCM}

\subsubsection*{Benefits of online model updates}
\begin{figure}[ht]
	\centering
	\includegraphics[width=0.4\textwidth]{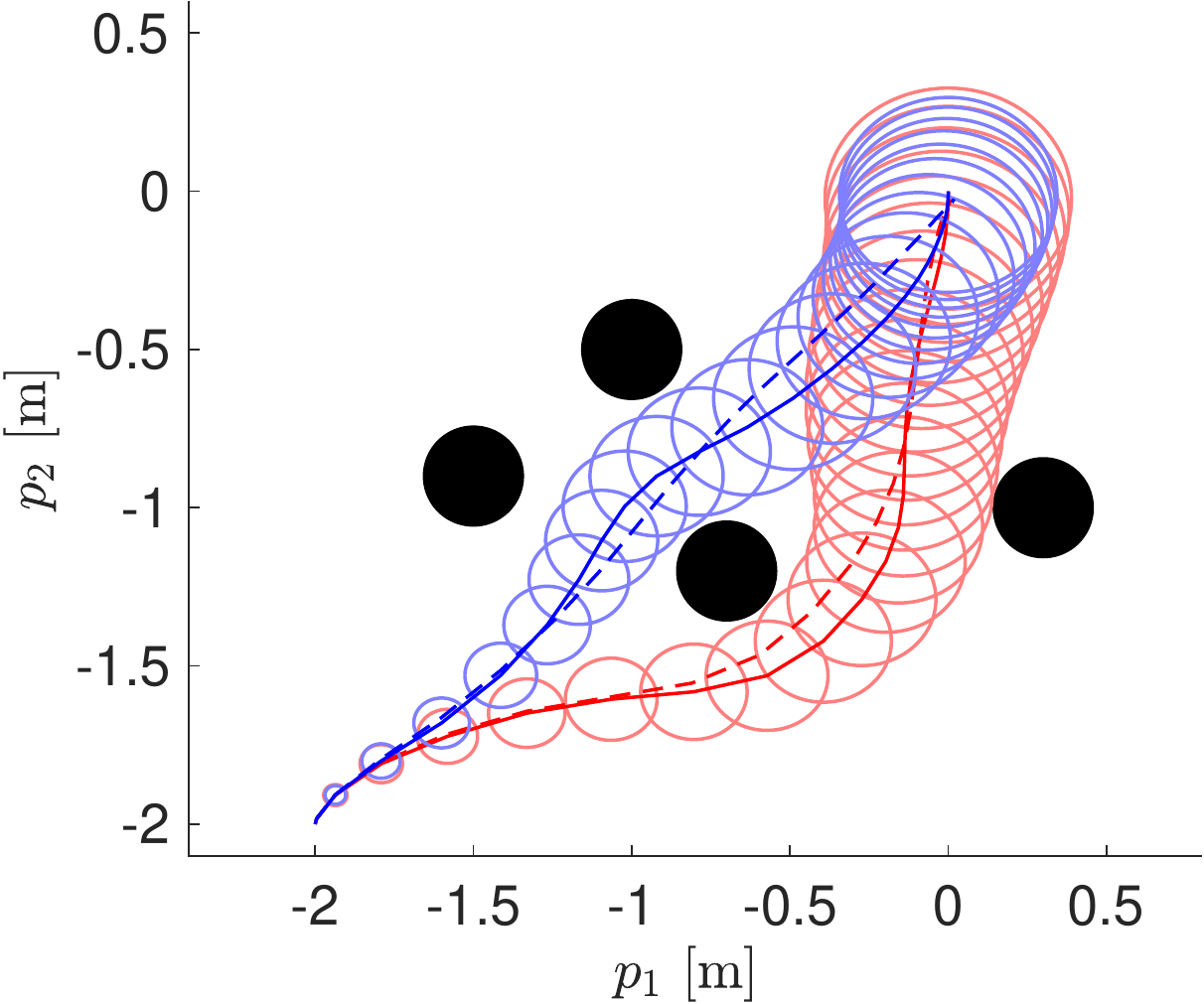}
	\caption{Comparison of the proposed RMPC scheme with (blue) and without (red) adaptation. The open-loop solutions at $t=T_\mathrm{s}$ are shown with nominal trajectories (solid) and ellipsoidal over-approximations of the tube based on $\underline{M}$. The closed-loop trajectories are dashed.} \label{fig:A-R}
\end{figure}

\begin{figure}[ht]
	\centering
	\includegraphics[width=0.4\textwidth]{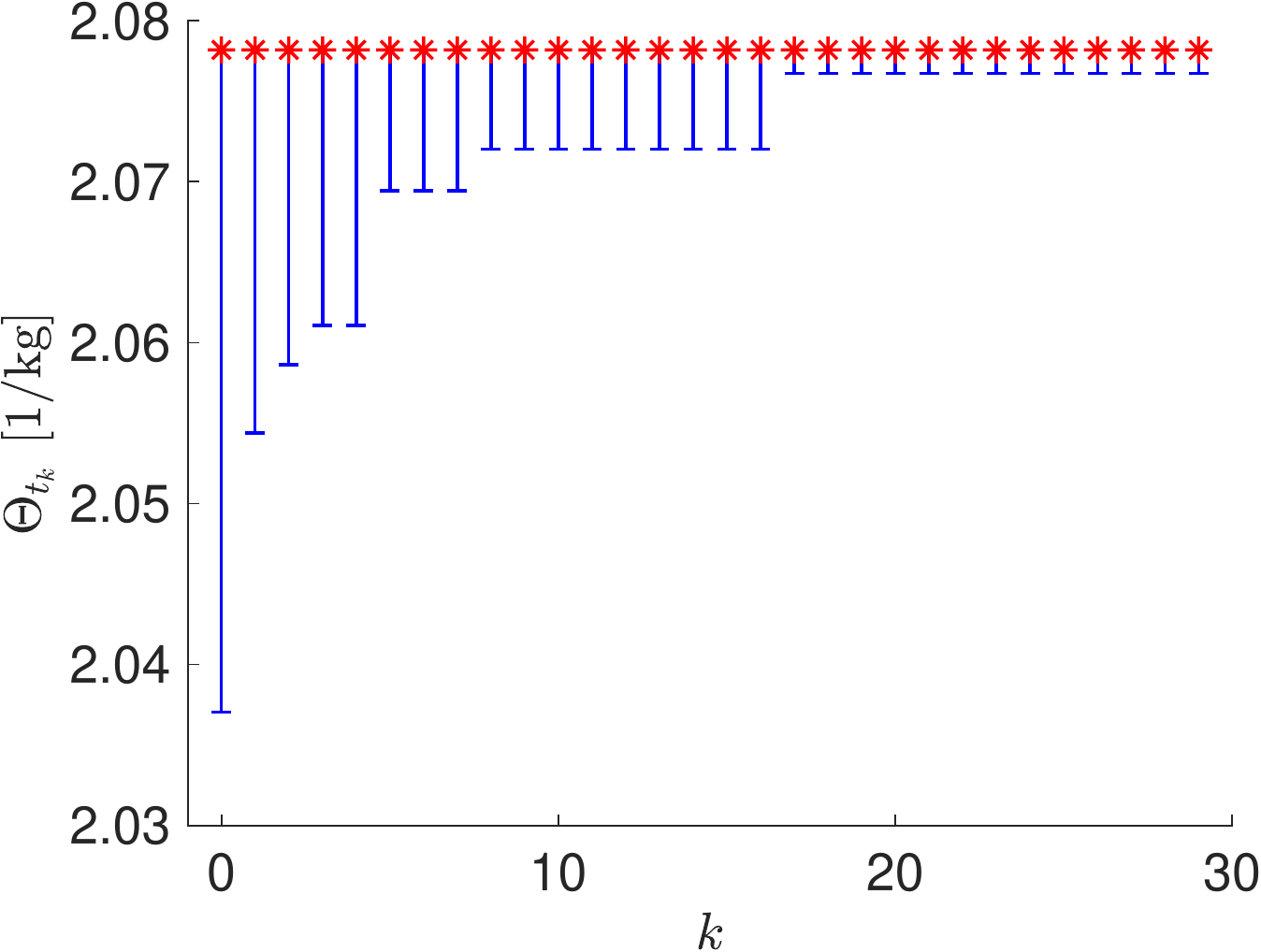}
	\caption{Evolution of the uncertain parameter set $\Theta_{t_k}$ (blue intervals) over the simulation steps $k$. The true parameter value is indicated by the red stars.}
	\label{fig:theta_evol}
\end{figure}

We demonstrate the benefits of online model updates using set-membership estimation (cf. Sec.~\ref{sec:set-update}) by comparing the proposed robust adaptive MPC to a purely robust MPC variation that uses $\Theta_{t_k}=\Theta_0$. The results are shown in Figure~\ref{fig:A-R}. Given the initial uncertainty, we cannot plan a safe trajectory through the smaller gap between the obstacles. Hence, both formulations plan a more conservative trajectory passing through the larger gap. After the first sampling period $t=T_\mathrm{s}$, we perform the set-membership update~\eqref{eq:Theta-update} using $n_m=1$, resulting in a $42\%$ reduction of the size of the parameter set $\Theta_{t_k}$. As a result,  the adaptive formulation can plan through the smaller gap.
As standard in (robust) MPC, the closed-loop trajectories can be significantly more aggressive than the conservative open-loop predictions due to the re-planning inherent in MPC. The evolution of the parameter set $\Theta_{t_k}$ in the simulation is depicted in Figure~\ref{fig:theta_evol}.

\subsubsection*{Comparison to rigid tube MPC}
\begin{figure}[ht]
	\centering
	\includegraphics[width=0.4\textwidth]{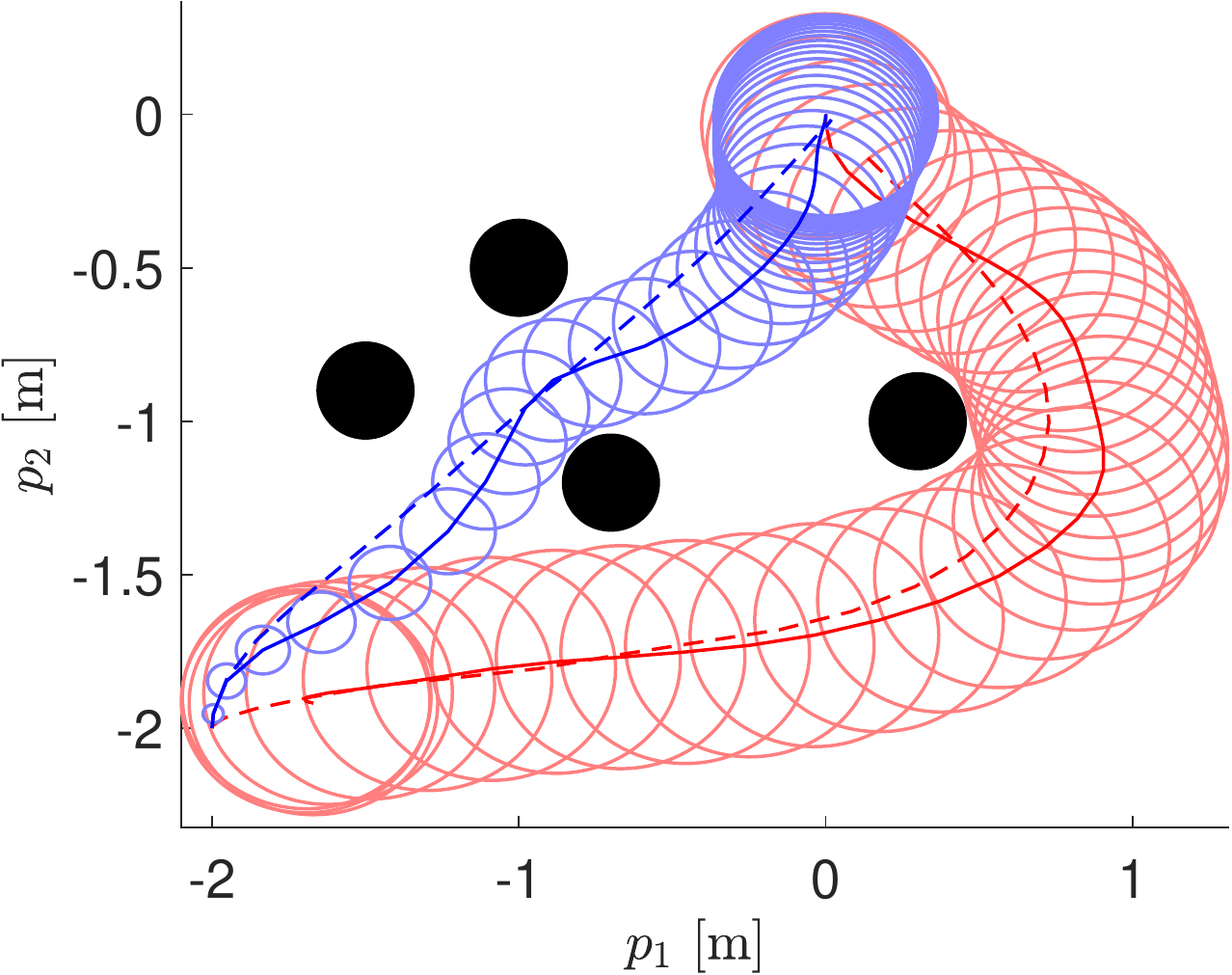}
    \caption{Comparison of rigid tube MPC (red) and the proposed homothetic tube RMPC (blue). The open-loop solutions at $t=0$ are shown with nominal trajectories (solid) and ellipsoidal over-approximations of the tube based on $\underline{M}$. The closed-loop trajectories are dashed.} \label{fig:R-CB}
\end{figure}

We also show the advantages of the state and input dependent disturbance bound~\eqref{eq:f-delta-def} in the proposed homothetic tube MPC scheme compared to a rigid tube approach, which is widely used in literature~\cite{mayne2005robust,yu2013tube,zhao2021tube,singh2017robust,singh2019robust}. To this end, we focus on the purely robust setting and use no online model updates. We compute a constant scaling $\overline{\delta}$ to recover the rigid tube, compare Section~\ref{sec:discussion}.
To ensure initial feasibility of this more conservative approach, we decrease the parametric uncertainty by 50\% and increase the prediction horizon to $N=35$ steps for both approaches. The resulting open-loop predictions and closed-loop trajectories can be seen in Figure~\ref{fig:R-CB}. 
The rigid tube (red) is significantly larger and correspondingly the planned trajectory cannot pass between the obstacles, but bypasses them from the right. 
On the other hand, the proposed homothetic tube MPC formulation finds a feasible solution through the smaller gap between the obstacles, demonstrating improved flexibility.

\subsubsection*{Computational complexity}
The offline computations to obtain the CCM and corresponding constants (Alg.~\ref{alg:offline}) took 10 minutes, while the just-in-time compilation (jit, cf.~\cite{andersson2019casadi}) of the RAMPC scheme  took another 20 minutes. 
The following table lists the online computation times for one iteration in IPOPT for the considered MPC schemes with a prediction horizon of $N=25$ steps. \\
\begin{tabular}{p{13mm}|p{13mm}|p{13mm}|p{12mm}|p{12mm}}
    RAMPC & RMPC & RMPC - Euler & Rigid tube & Nominal MPC  \\
    \hline
    60.46 ms & 54.64 ms & 19.52 ms & 8.79 ms & 4.32 ms\\
\end{tabular}

As one can see, adaptation leads to a relatively small increase in computation time (around 10\%) due to the additional variables $\overline{\theta}_t$, $\overline{\delta}_\mathrm{f}$ that appear in constraints \eqref{eq:MPC-dynamics}--\eqref{eq:MPC-theta-in-Theta}, \eqref{eq:MPC-terminal}. We also included an RMPC scheme that uses an explicit Euler discretization for the tube propagation~\eqref{eq:MPC-delta}. This simplification yields a significant reduction in computation times, yet the error in the tube propagation is below 5\% over the full horizon. 
Finally, the rigid tube approach has the lowest computational complexity among the RMPC frameworks, due to the absence of the tube dynamics~\eqref{eq:MPC-delta} leading to simpler constraints.\footnote{Among others, the tube dynamics~\eqref{eq:f-delta-def} require the expensive evaluation of $M(z)=W(z)^{-1}$. This inverse can also be avoided in the homothetic tube MPC by using the over-approximation $\overline{M}\succeq M(z)$ (cf. discussion Sec.~\ref{sec:discussion}).} 
Compared to a nominal MPC, the additional complexity of the rigid tube MPC is mainly due to the geodesic computation in the initial condition~\eqref{eq:MPC-delta0}.

\section{Conclusion}
\label{sec:conclusion}
We have presented a robust adaptive MPC framework that uses a homothetic tube based on general CCMs and incorporates recursive model updates using set-membership estimation. 
We demonstrated the improved flexibility using a nonlinear example.
Open-issues include the consideration of non-parametric model uncertainty and noisy output measurements.

\ack
This work has been supported by the Swiss National Science Foundation
under NCCR Automation (grant agreement 51NF40 180545)

\bibliographystyle{plain}        
\bibliography{Literature}

\appendix
\section{Appendix} \label{app:proof-V-delta}
In the following, we provide the proof of Proposition~\ref{prop:V-delta}, which builds on established arguments from contraction theory~\cite{lohmiller1998contraction,manchester2017control,singh2019robust}, more specifically also \cite[Thm. 1]{zhao2021tube} and \cite[App. A]{schiller2022lyapunov}.
\begin{pf}
	\textbf{Part I:} Condition \eqref{eq:CCM-bounded-M} together with the definition of $V_\delta$ \eqref{eq:V-delta-def} implies Inequality \eqref{eq:Vd-bound}, cf., \cite[Eq. (48)--(50)]{schiller2022lyapunov}. \\
	\textbf{Part II:} Recall that by assumption $(\gamma^\star(s), \gamma^{\mathrm{u}}(s)) \in \Z$ for all $s \in [0,1]$, and $\left.\dfrac{\mathrm{d} \gamma^\mathrm{u}}{\mathrm{d}s}\right|_s = K(\gamma^\star(s)) \gamma_{\mathrm{s}}^\star(s)$, cf., Equation \eqref{eq:kappa-def}. For any constant parameter $\overline{\theta} \in \Theta_0$, the differential dynamics along the geodesic are characterized by the Jacobian of the nonlinear system~\cite{zhao2021tube}, i.e.,
    \begin{align} 
        \dot{\gamma}^\star(s) =& f_\mathrm{w}(\gamma^\star(s),\gamma^\mathrm{u}(s),\overline{\theta} + s(\theta - \overline{\theta}),sd), \label{eq:app-gamma-dot} \\
        \dot{\gamma}_s^\star(s) =& A_\mathrm{cl}(\gamma^\star(s),\gamma^\mathrm{u}(s),\overline{\theta} + s(\theta - \overline{\theta}),sd) \gamma_{\mathrm{s}}^\star(s) \label{eq:app-gamma-dot-s} \\
        &+ \gamma^\mathrm{w}(s), \nonumber \\
        \gamma^\mathrm{w}(s) :=& G(\gamma^\star(s),\gamma^\mathrm{u}(s))(\theta - \overline{\theta}) + E(\gamma^\star(s))d. \label{eq:app-gamma_w}
    \end{align}
    Note that $\overline{\theta} + s(\theta - \overline{\theta}) \in \Theta_0$ and $sd \in \D$, $s \in [0,1]$, since $0\in\D$, and $\Theta_0$, $\D$ are convex. Let us define the following abbreviations:
    \begin{align*}
        \hat{G}(s):=&M(\gamma^\star(s))^{\frac{1}{2}}G(\gamma^\star(s),\gamma^\mathrm{u}(s)), \\
        \hat{E}(s):=&M(\gamma^\star(s))^{\frac{1}{2}}E(\gamma^\star(s)), \\
        G_{\mathrm{s},k}(z,v):= & \left. \dfrac{\partial \left(M^{\frac{1}{2}}[G]_{:,k}\right)}{\partial x}\right|_{(z,v)}\\
        &+ \left.\dfrac{\partial \left(M^{\frac{1}{2}}[G]_{:,k}\right)}{\partial u}\right|_{(z,v)}K(z), \\
        E_{\mathrm{s},j}(z) := & \left.\dfrac{\partial \left(M^{\frac{1}{2}}[E]_{:,j}\right)}{\partial x}\right|_{z},\quad k\in\mathbb{I}_{[1,p]},\quad j\in\mathbb{I}_{[1,q]}.
    \end{align*}
    Note that for any $\tilde{\theta} \in \R^p$ and any $d\in \R^q$, we have 
    \begin{align*}
        \left.\dfrac{\mathrm{d} \hat{G}}{\mathrm{d} s}\right|_s \tilde{\theta} =& \sum_{k=1}^p [\tilde{\theta}]_k G_{\mathrm{s},k}(\gamma^\star(s),\gamma^\mathrm{u}(s)) \gamma_{\mathrm{s}}^\star(s), \\
        \left.\dfrac{\mathrm{d} \hat{E}}{\mathrm{d} s}\right|_s d =& \sum_{j=1}^q [d]_j E_{\mathrm{s},j}(\gamma^\star(s)) \gamma_{\mathrm{s}}^\star(s).
    \end{align*}
    Hence, for any $s\in[0,1]$ it holds:
    \begin{align} 
        &\|(\hat{G}(s) - \hat{G}(0))(\theta - \overline{\theta})\| \leq \int_0^s \left \|\left.\dfrac{\mathrm{d} \hat{G}}{\mathrm{d} s}\right|_{\tilde{s}} (\theta -\overline{\theta})\right \| \mathrm{d}\tilde{s} \nonumber \\
        =& \int_0^s \left \|\sum_{k=1}^p G_{\mathrm{s},k}(\gamma^\star(\tilde{s}), \gamma^\mathrm{u}(\tilde{s}))[\theta - \overline{\theta}]_k \gamma_{\mathrm{s}}^\star(\tilde{s}) \right \| \mathrm{d}\tilde{s} \nonumber \\
        \leq & \sum_{k=1}^p |[\theta - \overline{\theta}]_k| \int_0^s \left \|G_{\mathrm{s},k}(\gamma^\star(\tilde{s}), \gamma^\mathrm{u}(\tilde{s})) M(\gamma^\star(\tilde{s}))^{-\frac{1}{2}} \right \|  \nonumber\\
        &\cdot \left \|M(\gamma^\star(\tilde{s}))^{\frac{1}{2}} \gamma_{\mathrm{s}}^\star(\tilde{s}) \right \| \mathrm{d} \tilde{s}  \nonumber \\
        \stackrel{\eqref{eq:V-delta-def}}{\leq} & \sum_{k=1}^p L_{\mathrm{G},k}|[\theta - \overline{\theta}]_k| V_\delta(x,z), \label{eq:app-L-G-deriv}
    \end{align}
    with 
    \begin{align} \label{eq:app-L-G}
        L_{\mathrm{G},k} := \max_{(z,v) \in \Z} \left \| G_{\mathrm{s},k}(z,v) M(z)^{-\frac{1}{2}} \right \|, \; \; k\in\I{1,p}.
    \end{align}
    Similarly, for any $s \in [0,1]$, $d \in \D$, we get   
    \begin{align} \label{eq:app-L-D-deriv}
        \left \|  (\hat{E}(s) - \hat{E}(0) )d \right \| \leq L_\D V_\delta(x,z),
    \end{align}
    with
    \begin{align} \label{eq:app-L-D}
        L_\D := \max_{z \in \Z_\mathrm{x}, d \in \D} \left \| \sum_{j=1}^q E_{\mathrm{s},j}(z) M(z)^{-\frac{1}{2}} [d]_j \right \|.
    \end{align}
    The above inequalities yield
    \begin{align} 
        &\int_0^1 \gamma_{\mathrm{s}}^\star(s)^\top M(\gamma^\star(s))\gamma^\mathrm{w}(s) \mathrm{d}s \nonumber \\
        \leq & \int_0^1 \left \|  \gamma^\star_s(s)^\top \left (M(\gamma^\star(s))^{\frac{1}{2}} \right)^\top \right \| \left( \left \| M(\gamma^\star(0))^{\frac{1}{2}} \gamma^\mathrm{w}(0) \right \| \right. \nonumber \\
        & \left. + \left \|M(\gamma^\star(s))^{\frac{1}{2}} \gamma^\mathrm{w}(s) -  M(\gamma^\star(0))^{\frac{1}{2}}\gamma^\mathrm{w}(0) \right \|\right) \mathrm{d}s \nonumber \\
        \stackrel{\substack{\eqref{eq:V-delta-def} \\ \eqref{eq:app-gamma_w}}}{\leq} & V_\delta(x,z) \left (
        \vphantom{+ \max_{s \in [0,1]}  \left \|M(\gamma^\star(s))^{\frac{1}{2}} \gamma^\mathrm{w}(s) -  M(\gamma^\star(0))^{\frac{1}{2}}\gamma^\mathrm{w}(0) \right \|} \|G(z,v)(\theta - \overline{\theta}) + E(z)d \|_{M(z)} \right. \nonumber \\
        & \left. + \max_{s \in [0,1]}  \left \|M(\gamma^\star(s))^{\frac{1}{2}} \gamma^\mathrm{w}(s) -  M(\gamma^\star(0))^{\frac{1}{2}}\gamma^\mathrm{w}(0) \right \| \right) \nonumber \\
        \stackrel{\substack{\eqref{eq:app-L-G-deriv} \\ \eqref{eq:app-L-D-deriv}}}{\leq} & \left\|G(z,v)(\theta - \overline{\theta}) + E(z)d \right\|_{M(z)} V_\delta(x,z) \nonumber \\
        &+ \left(L_\D + \sum_{k=1}^p L_{\mathrm{G},k} |[\theta - \overline{\theta}]_k|\right) V_\delta(x,z)^2. \label{eq:app-int-distbound}
    \end{align}
    The Riemannian energy defined as\\ $\mathcal{E}(x,z) := \int_0^1\gamma_{\mathrm{s}}^\star(s)^\top M(\gamma^\star(s)) \gamma_{\mathrm{s}}^\star(s) \mathrm{d}s$~\cite{manchester2017control,singh2019robust,zhao2021tube} satisfies
    \begin{align}
        \dot{\mathcal{E}}(x,z) = &\int_0^1 \gamma_{\mathrm{s}}^\star(s)^\top \dot{M}(\gamma^\star(s)) \gamma_{\mathrm{s}}^\star(s) \nonumber \\
        & + \dot{\gamma}_s^\star(s)^\top M(\gamma^\star(s)) \gamma_{\mathrm{s}}^\star(s) \nonumber \\
        & + \gamma_{\mathrm{s}}^\star(s)^\top M(\gamma^\star(s)) \dot{\gamma}_s^\star(s) \mathrm{d}s \nonumber \\
        \stackrel{\substack{\eqref{eq:app-gamma-dot} \eqref{eq:app-gamma-dot-s} \nonumber \\ \eqref{eq:CCM-contraction}}}{\leq} & \int_0^1 -2\rho_\mathrm{c} \gamma_{\mathrm{s}}^\star(s)^\top M(\gamma^\star(s)) \gamma_{\mathrm{s}}^\star(s) \nonumber \\
        & + 2\gamma_{\mathrm{s}}^\star(s)^\top M(\gamma^\star(s))\gamma^\mathrm{w}(s) \mathrm{d}s \nonumber  \\
        \stackrel{\eqref{eq:app-int-distbound}}{\leq}& -2\rho_\mathrm{c}\mathcal{E}(x,z) +2V_\delta(x,z) \nonumber \\
        & \cdot \left( \left(L_\D + \sum_{k=1}^p L_{\mathrm{G},k} |[\theta - \overline{\theta}]_k|\right) V_\delta(x,z) \right. \nonumber \\
        & \left.+ \left\|G(z,v)(\theta - \overline{\theta}) + E(z)d \right\|_{M(z)} \vphantom{\sum_{k=1}^p}\right). \label{eq:app-energy}
    \end{align}
    Furthermore, since $\mathcal{E}(x,z) = V_\delta(x,z)^2$ (see \cite{manchester2017control,singh2017robust}), it holds that $\dot{\mathcal{E}}(x,z) = 2V_\delta(x,z)\dot{V_\delta}(x,z)$, and hence, Inequality \eqref{eq:app-energy} is equivalent to Condition~\eqref{eq:Vd-contraction}. $~\hfill \square$\\
\end{pf}
\end{document}